\documentclass[3p, 11pt]{elsarticle}
\usepackage{graphicx,amscd,amsmath,amssymb,verbatim}
\usepackage{amsfonts,epsfig}
\usepackage{mathptmx}
\usepackage{setspace}
\usepackage{epsfig}
\usepackage{url}
\usepackage{multirow}
\usepackage{color}

\begin{document}

\journal{Elsevier}

\begin{frontmatter}

\title{Universality, Correlations, and Rankings in the Brazilian Universities National Admission Examinations}

\author{Roberto da Silva$^1$, Luis C. Lamb$^2$, Marcia C. Barbosa$^1$} 

\address{1-Instituto de Fisica, Universidade Federal do Rio Grande do Sul,
Av. Bento Gon\c{c}alves, 9500 - CEP 91501-970, Porto Alegre, Rio Grande do
Sul, Brazil
{\normalsize{E-mail:{rdasilva,marcia.barbosa}@if.ufrgs.br}}}

\address{2-Instituto de Informatica, Universidade Federal do Rio Grande do Sul,
Av. Bento Gon\c{c}alves, 9500 - CEP 91501-970, Porto Alegre, Rio Grande do
Sul, Brazil
{\normalsize{E-mail:lamb@inf.ufrgs.br}}}

\begin{abstract}
The scores obtained by students
that have performed the  ENEM exam, the Brazilian High School National  Examination used  to admit students at the Brazilian universities, is analyzed. 
The average high school's scores 
are compared between different disciplines through 
the Pearson correlation coefficient. The results show
a very large correlation between the performance in  the
different subjects. 
Even thought the students' scores in the ENEM due
to the standardization form 
a Gaussian, we show that
the high schools' scores form a  bimodal distribution
that can not be used to evaluate and compare performance over time.
We also 
show that this high schools distribution reflects the  correlation
between school performance and economic level of the students.
The ENEM's scores are compared with a Brazilian non standardized exam,
the entrance exam at the Universidade Federal do Rio Grande do Sul.
The comparison of the performance of the same individuals in
both tests is compared showing that the two tests not
only select different abilities but chooses a different set
of individuals.
Our results indicates that standardized exams might be 
an interesting tool to compare performance over the years
but only of individuals and not of institutions. 
      
\end{abstract}

\end{frontmatter}

\setlength{\baselineskip}{0.7cm}

\section{Introduction}

\label{Section:Introduction} The selection of the part of the population
that have access to high education is challenge particularly because this
has important implication in the future of nation. China was the first
country in the world to face this challenge. The imperial examination
created in 605 during the Sui Dynasty was a civil service examination system
in Imperial China to select candidates for the state bureaucracy. This
system persisted until its extinction in 1905~\cite{Eb10}.

Recognizing that having a standardized test to select the elite would
guarantee the future of the United Kingdom, the idea of the test was
introduced into Europe in the early 19th century by the Britain's consul in
Guangzhou, China, Thomas Taylor Meadows~\cite{Hu96}. In 1806 the United
Kingdom introduced the selection of public servants through an examination.

In the high education system the standardized test was first employed by
Napoleon that created le baccalaur\'{e}at or simply le bac. In the United
Kingdom it was created the the General Certificate of Secondary Education.
It was from Britain that standardized testing spread, not only throughout
the British Commonwealth, but to Europe and then America. In the United
States two systems dominate the selection of the universities: the
Scholastic Aptitude Test (SAT) and the American College Testing (ACT)
created in 1926 and in 1959 respectively. The first focus on evaluation
abilities while the second measures deduction skills.

The current standardize tests in the United States, European Countries and
Asia have in common that they are organized in such a way that the scores
follow a normal distribution~\cite{Do02,Ko00}, $f(x)$, that is characterized
by the mean $\left\langle x\right\rangle $ and standard deviation $\sigma$.
The result of a particular candidate in one test, $x_{i}$, becomes
universally comparable by the regular transformation $z_{i}=\left(
x_{i}-\left\langle x\right\rangle \right) /\sigma $.

Even thought quite appealing due to its simplicity, the use of standardized
tests to select the entrance at the universities is not free from criticisms~%
\cite{Co04,Ya07}. Because the exams are tested in a biased population,
minorities and foreigners show difficulties in understanding the cultural
subtleties~\cite{Cu09,Ko13}. In addition it is not clear that the of one
year or one test can be compared with the results from other years or other
tests simply by performing a good performance in the college is correlated
with the scores obtained at the standardized tests but with the performance
at the high school~\cite{Ti01}.

In the particular case of the United States, since the admission is a
multidimensional process in which not only the SAT or ACT scores, but also
the performance at the high school, recommendation letters and extra
curricular activities are taken into account; the criticisms to the
standardized test method imply a lower impact in the selection process when
compared with countries in which the score is the only evaluated dimension.
In addition, other countries have a number of competing standardized tests
what also guarantee that the education does not become hostage of one
evaluation method.

A proper analysis of the standardized tests that would answer to the
criticisms~\cite{Ti01} to the method is not possible, since the scores of
all these standardized exams are not available for detailed analysis.

In Brazil the procedure to enter at the high education system up to the end
of the 20th century were exams organized by each college. Even though this
method guarantee diversity in the selection process, it made mobility of the
student rather difficult. In addition, differently from the United States
and some European Union countries, Brazil does not use an university
admission system based on historical or annual tests of high school students
but only this entrance exam. Typically, the university entrance examination
is composed of several multiple choice exams which encompass all high school
subject areas.

Over the last decade, the Brazilian government has introduced a standardized
university entrance examination known as ENEM (\emph{pt}: Exame Nacional do
Ensino M\'{e}dio - \textit{en}: High School National Examination). This new
exam is applied across the country what allows for mobility of the students
from one state to the other, uses a methodology that allow for comparison of
the scores obtained in one year with the scores of the previous years and is
elaborated in a centralized form. The major drawback of using one unique
exam to select the students is that the system becomes dependent of type of
analysis. Additional problems are the following. The exam is too ample. It
covers a very large number of questions, and many students are not able to
finish the (long) exam questions in the allowed exam time. This means that
questions are not homogeneously solved by the students since they possibly
solve the questions in different samples. Thus, candidates with partial
knowledge of the high school subjects potentially can have the same
opportunity (and perform similarly to) a candidate with a comprehensive
background. Extensive, unclear and redundant question statements take too
much time to read and grasp and do not explore relevant knowledge of the
students; rather, understanding a question statement has affected the
students performance. \qquad \qquad \qquad\ \ 

Moreover, it is also important to mention the unclear methods used to
calculate the examination scores and the absence (or lack of) brute scores
for external analyses by the independent scientific community. Finally, no
changes have been made in the exam methodology since its inception, which
could lead to improvements in the test questions.

Despite the many criticisms about the contents of the ENEM's questions~\cite%
{La14,La15}, the process has its merits. If it is managed and carried out
properly it would lead to an interesting mechanism to tackle biases and
distortions towards bringing a larger contingent of state owned high school
students to public universities.

However, before it becomes a unique tool to evaluate all the students in
Brazil, the ENEM has to be evaluated and tested against another existing
local exams. As far as we know this was never done with the other
standardized tests, maybe with the exception of the SAT which performance
has been checked against high school grades but for a very narrow number of
students~\cite{Ti01}.

In this paper statistical physics tools are employed to understand the
universal aspects of this exam. This strategy is not new and has been used
to analyze high school performance~\cite{Gligor2003,Leonard1992,Neelon2014}.
The scores that the students obtained in the different disciplines in the
ENEM are analyzed. In addition the scores obtained by the different high
schools in the same exam are also evaluated. Finally a comparison between
the performance of a selected number of students at the ENEM and at a local
exam at one specific university, the Universidade Federal do Rio Grande do
Sul (UFRGS), during three consecutive years is also shown providing a unique
tool to identify what differ in the profile of the students selected by both
methods.

The remaining of the paper goes as follows. In the sec. \ref{Section:Data}
the data set employed in this work is introduced, in the sec. \ref%
{Section:Results} the results are presented and conclusions summarize the
paper in sec. \ref{Section:Conclusions}.

\section{Data Set Analysis}

\label{Section:Data}

The first part of our data set supplies the average scores of the $14,715$
high schools from Brazil in 2013 considering: School percentage
(participation) rate of their students and the economic (average family
income) level of the school.

The exam is composed by tests in five different school subject: Writing,
Language, Human Sciences, Natural Sciences, and Mathematics. The economic
(income) level of the schools are divided into 7 different levels: very
high, high, high average, average, low average, low, and very low. We
attributed $3,2,1,0,-1,-2,-3$ respectively for these levels.

The second data set supplies ENEM and UFRGS entrance examination scores of
the students that have taken both exams. We have analyzed students by three
consecutive years 2011, 2012 and 2013. Here we have cleaned the data by
extracting students that have score zero in one or more school subjects. For
example for 2011 we after cleaning we have $11,515$\ students that performed
the exam the school subject Writing at UFRGS. From these students only $%
10,315$\ had also non zero score at this same school subject at UFRGS, which
is the minimum (worst case) size sample used in our work for all comparison
tests (Pearson correlation and ranking tests) used in this work. This means
that in all possible cross over between two school subjects considering all
combinations: UFRGS with UFRGS or ENEM with ENEM or even UFRGS with ENEM we
had always larger samples. This guarantees the good significance in our
calculations. For example we find in 2013 more than $25,000$\ students that
performed the school subject math in both UFRGS and ENEM with non zero score.

The high school subjects of the UFRGS entrance examination that we
considered compatible for a suitable comparison with ENEM are: Writing,
Geography, History, Physics, Chemistry, Biology, and Mathematics. Writing
and Mathematics have a direct association between the UFRGS and ENEM
examinations. For our purposes, we associate Humanities (ENEM) with
Geography and History (UFRGS), and Natural Sciences (ENEM) with Physics,
Chemistry, and Biology (UFRGS).

\section{Results}

\label{Section:Results}

\subsection{ENEM Scores in the Brazilian High Schools}

\begin{figure}[th]
\begin{center}
\includegraphics[width=0.35\columnwidth]{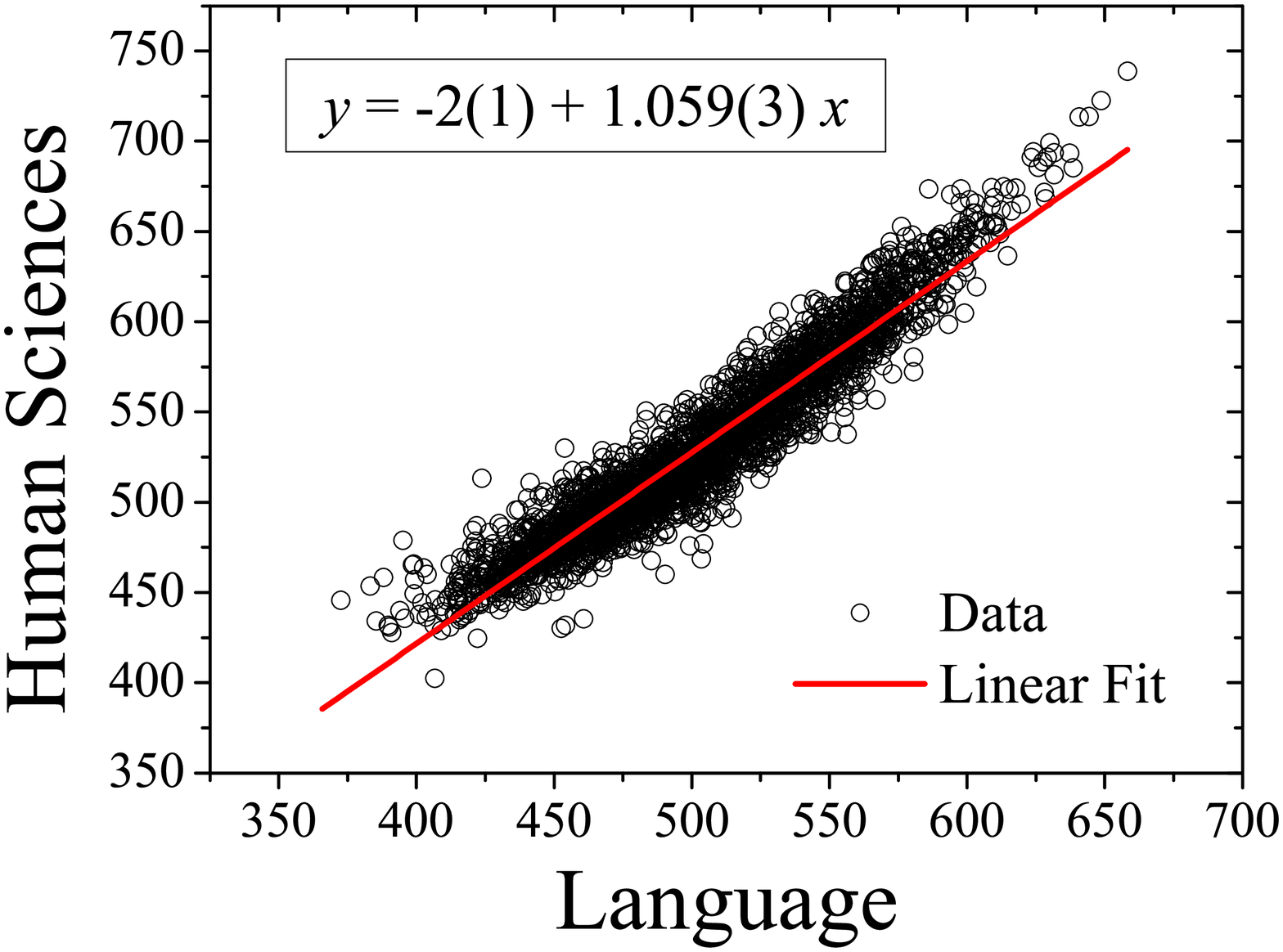}%
\includegraphics[width=0.35\columnwidth]{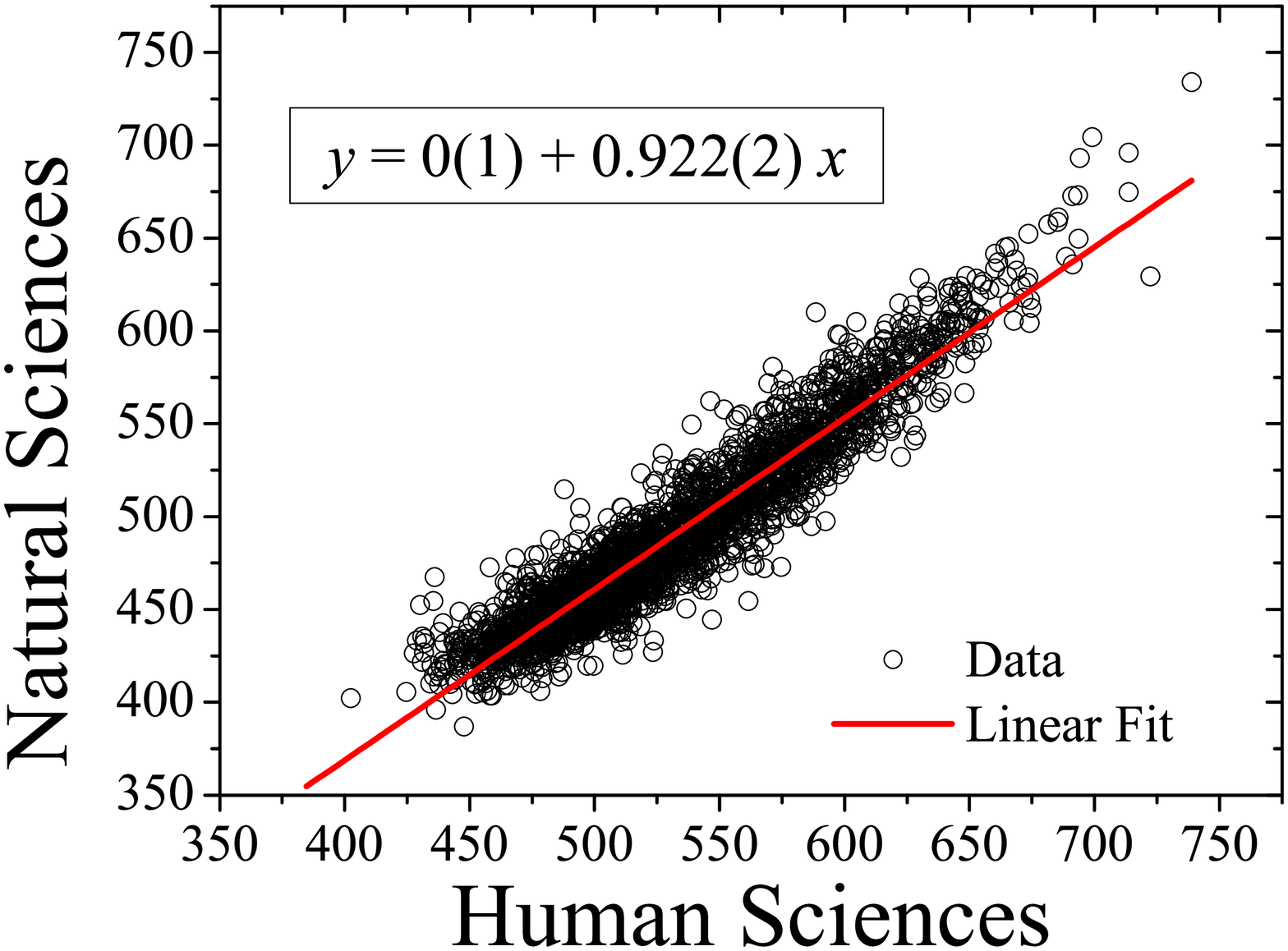} %
\includegraphics[width=0.35\columnwidth]{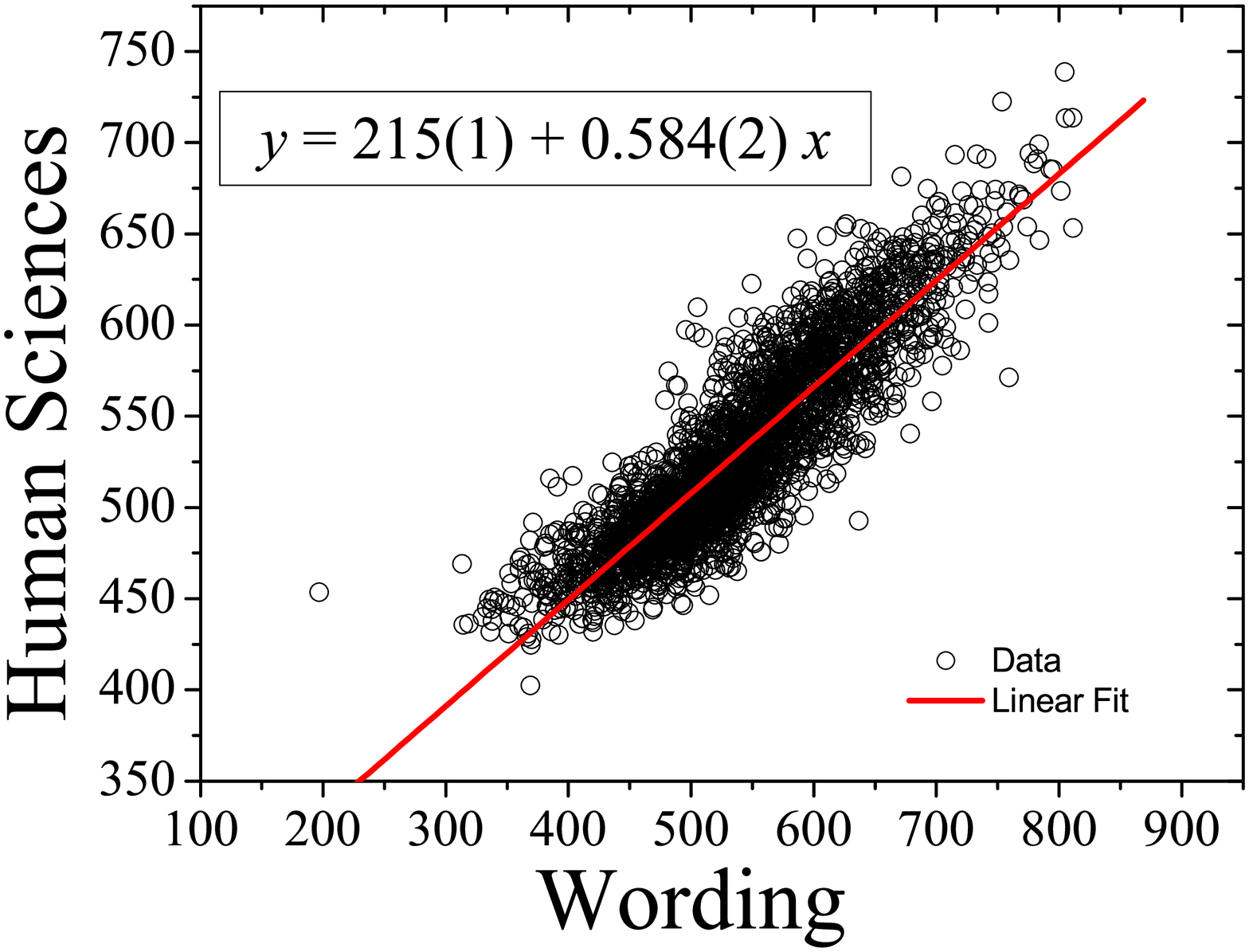}%
\includegraphics[width=0.35\columnwidth]{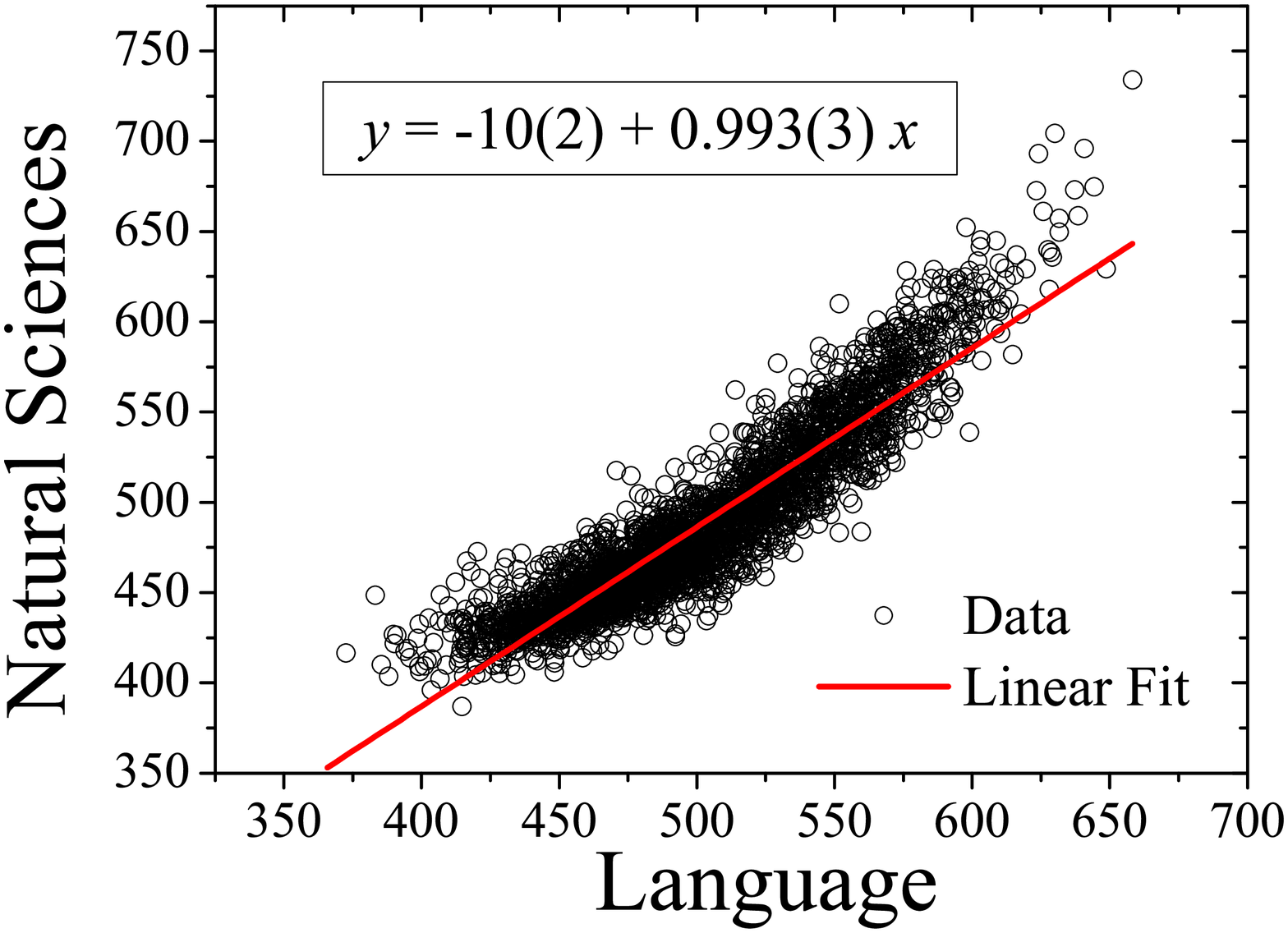} %
\includegraphics[width=0.35\columnwidth]{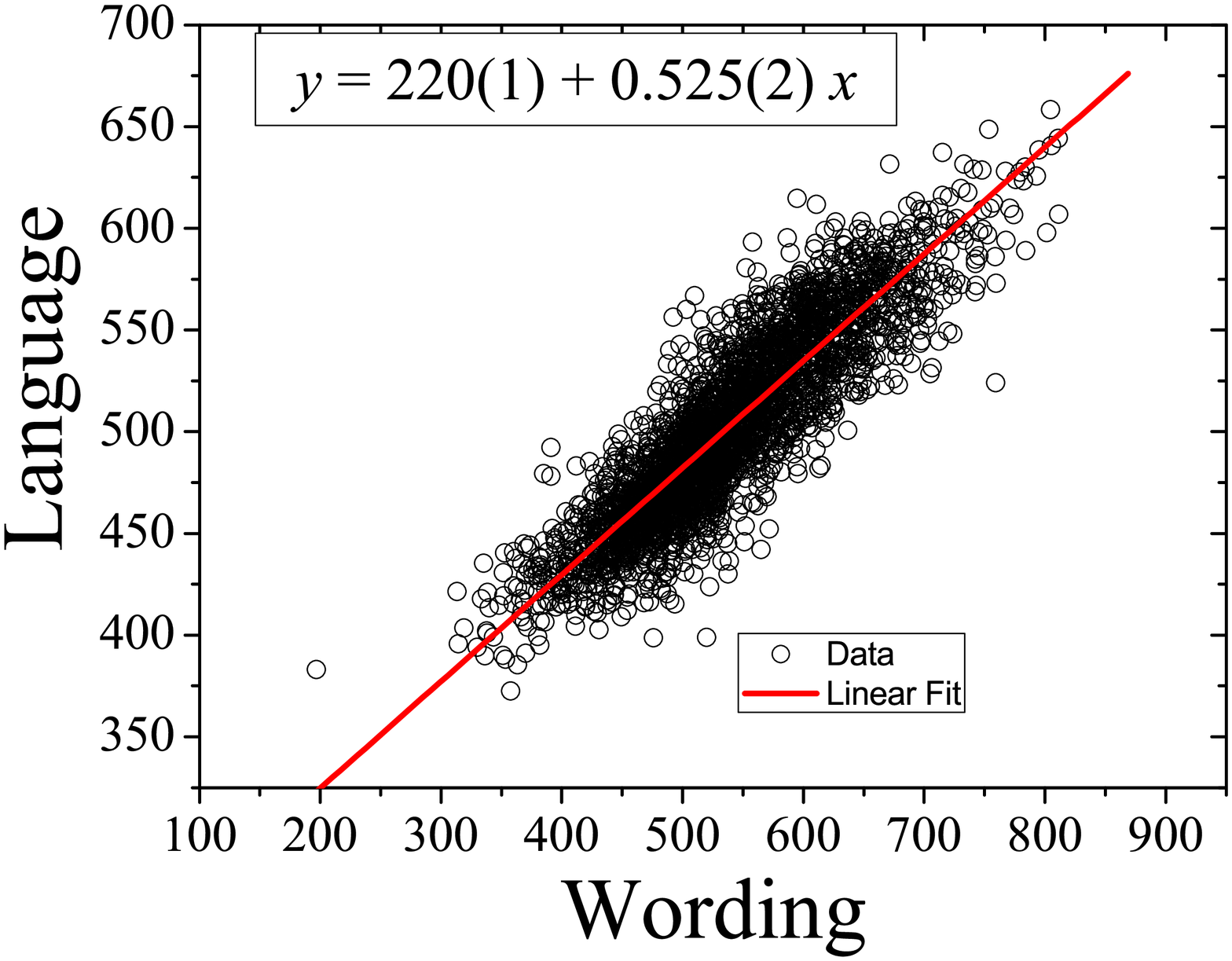}%
\includegraphics[width=0.35\columnwidth]{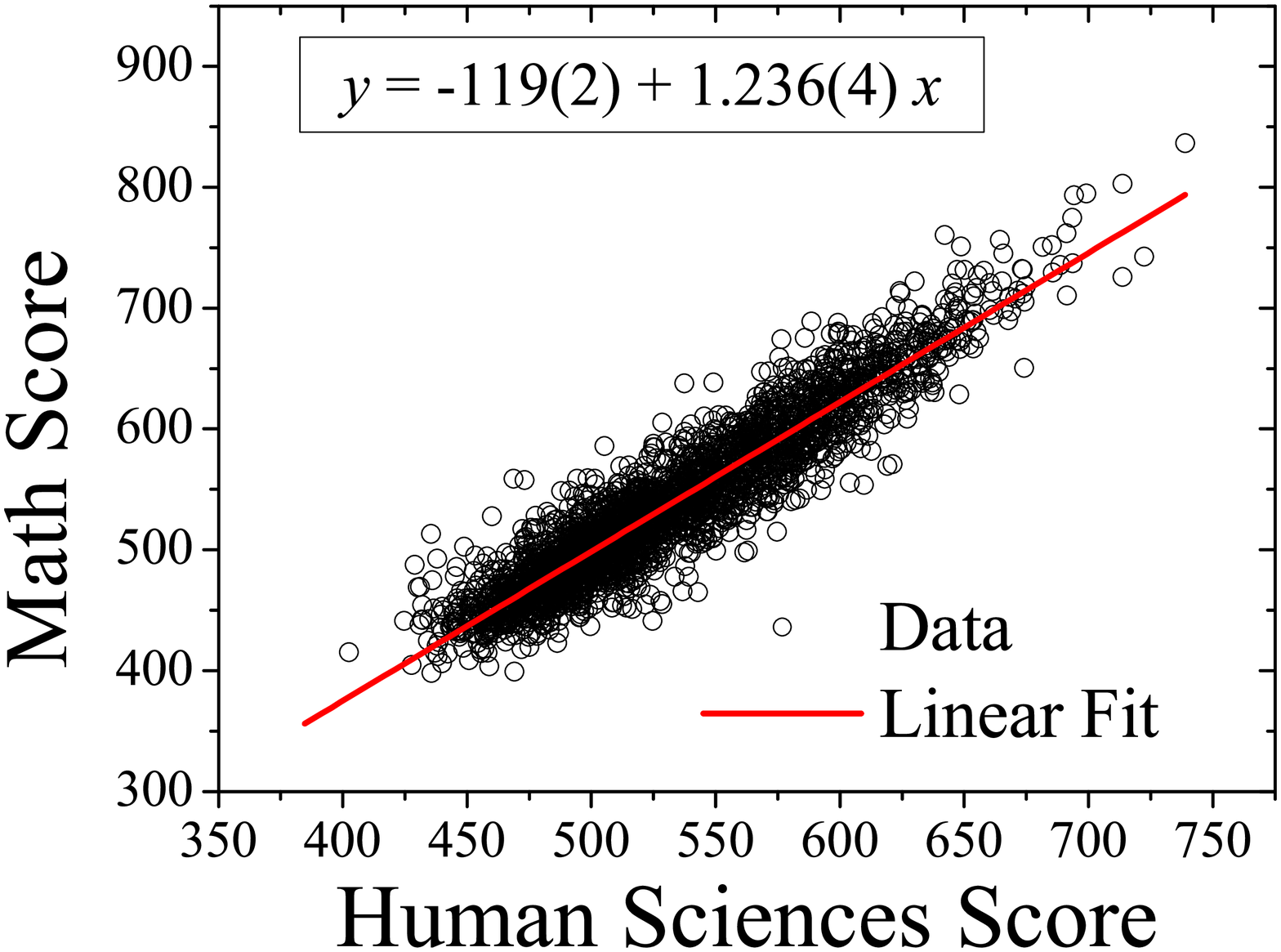} %
\includegraphics[width=0.35\columnwidth]{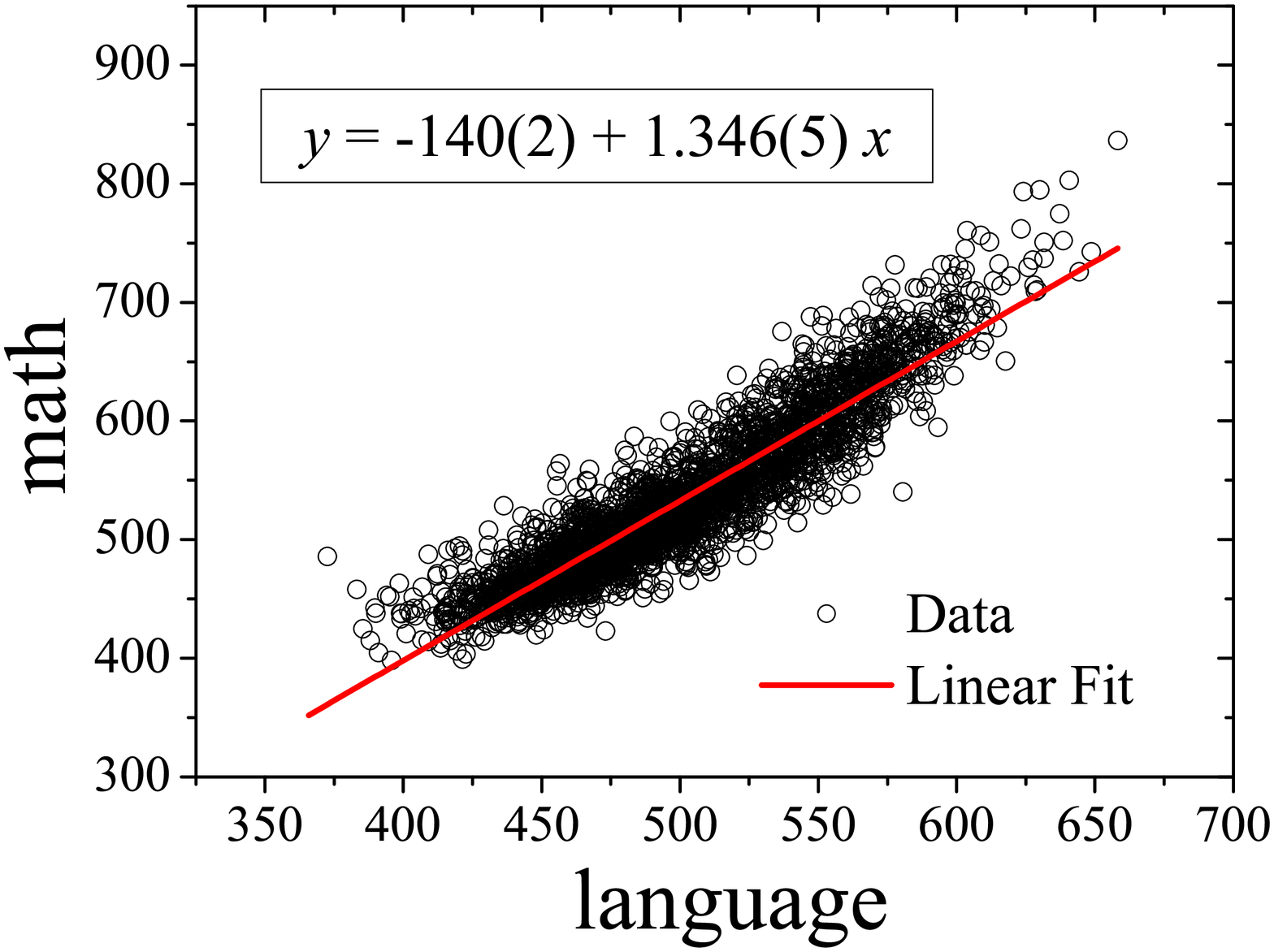}%
\includegraphics[width=0.35\columnwidth]{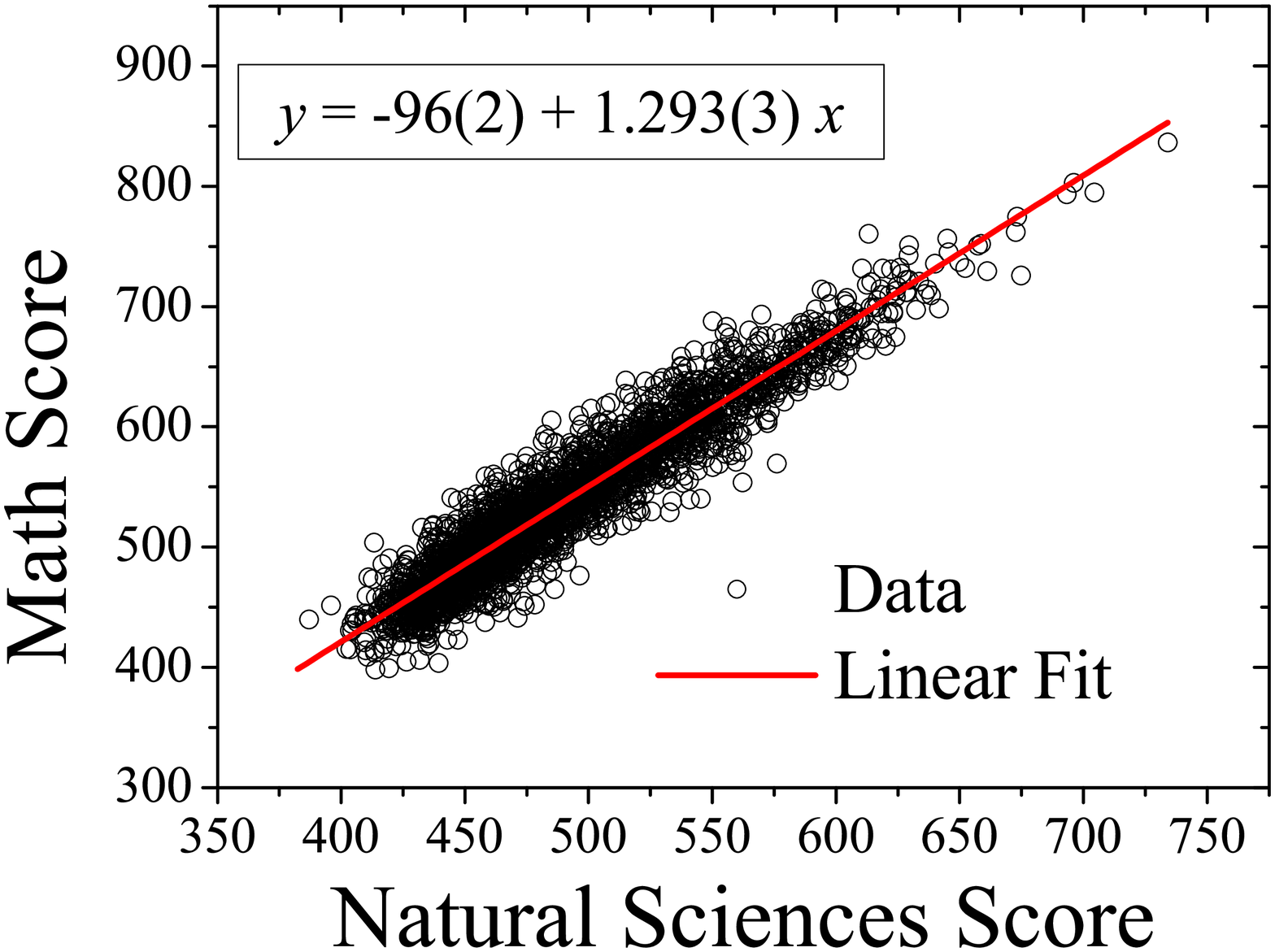}
\end{center}
\caption{Scattering diagrams for different pairs of high school subjects.
Visually, we can observe a good correlation (pairwise) between them. }
\label{correlation_scattering}
\end{figure}

First, the correlations between the scores at different subjects obtained by
all the high schools were computed. The Fig.~\ref{correlation_scattering}
illustrates the comparison between these scores . Visually, these diagrams
show a strong linear correlation between the scores of different subjects.
This indication can be quantified by a single number,the Pearson correlation
coefficient given by%
\begin{equation}
r=\frac{\sum_{i=1}^{n}(x_{i}-\left\langle x\right\rangle
)(y_{i}-\left\langle y\right\rangle )}{\sqrt{\sum_{i=1}^{n}(x_{i}-\left%
\langle x\right\rangle )^{2}}\sqrt{\sum_{i=1}^{n}(y_{i}-\left\langle
y\right\rangle )^{2}}}  \label{Eq.Pearson}
\end{equation}%
where $x_{i}$ and $y_{i}$ represent the scores of two different subjects
associated to $i$-th institution, with $i=1,...,n$. The values of $r$ vary
from $-1$ when the two data sets are negatively correlated, to $0$ when they
are uncorrelated up to $1$ when they are positively correlated. Since $r$ is
computed averaged over all the $n=14,715$ schools it gives a robust
indication of the correlations between the performance of the schools in the
different topics.

Table~\ref{Table:correlations} illustrates the values of $r$\ for the
different pair of topics. We can observe a high correlation among the
different schools which is not a surprise indeed, since the score schools
are more representative because represent averages over many students.
However some particularities must be mentioned. All subjects are more
correlated with Language and Humanities (or Human Sciences) than with
Writing. This is quite surprising since in principle one would expect that
Natural sciences and Mathematics would show a less evident correlation with
Language or Humanities. Language and Human Sciences are slightly more
correlated with writing than Natural Sciences and Maths. Although the
biggest correlations are found in the somewhat more intuitive cases: between
Language and Human Sciences and between Natural Sciences and Maths, $%
r=0.9554 $ and $r=0.9531$ respectively; we also found $r=0.9523$ between
Human and Natural Sciences and $r=0.9408$ between Human Sciences and Maths,
which are not expected results if the analysis was made with correlation for
the different schools. The last row of the table (in bold) corresponds to
correlation coefficients between each school subject and the average final
score of the schools what is quite strong. This indicates that either the
schools in Brazil shown not specific strength in any subject or the exam is
unable to capture the difference in the performance of the schools in
different areas of knowledge.

\begin{table}[tbp] \centering%
\begin{tabular}{ccccc}
\hline\hline
Writing & Language & Human Sciences & Natural Sciences & Math \\ \hline\hline
Writing & $0.8878$ & $0.8899$ & $0.8624$ & $0.8555$ \\ 
-- & Language & $0.9554$ & $0.9250$ & $0.9243$ \\ 
-- & -- & Human Sciences & $0.9523$ & $0.9408$ \\ 
-- & -- & -- & Natural Sciences & $0.9531$ \\ \hline
$\mathbf{0.8918}$ & $\mathbf{0.9694}$ & $\mathbf{0.9823}$ & $\mathbf{0.9791}$
& $\mathbf{0.9801}$ \\ \hline\hline
\end{tabular}%
\caption{Pearson correlation coefficients, $r$, between the two subjects  
scores in the ENEM 2013. The last row corresponds to the coefficient 
between each school subject score and the average score of the institution.}%
\label{Table:correlations}%
\end{table}%

One of the promises of performing a standardized exam is that it would make
possible for students coming from disadvantaged areas and schools to enter
at the major universities of the country. In order to test this hypothesis,
two different parameter were compute: (a) the scores as a function of the
social-economic level of the schools and (b) the score as a function of the
number of the students' participation at ENEM, namely the ratio between the
number of students that effectively took the examination and the total
number of students that were eligible to take the examination.

\begin{figure}[th]
\begin{center}
\includegraphics[width=0.75\columnwidth]{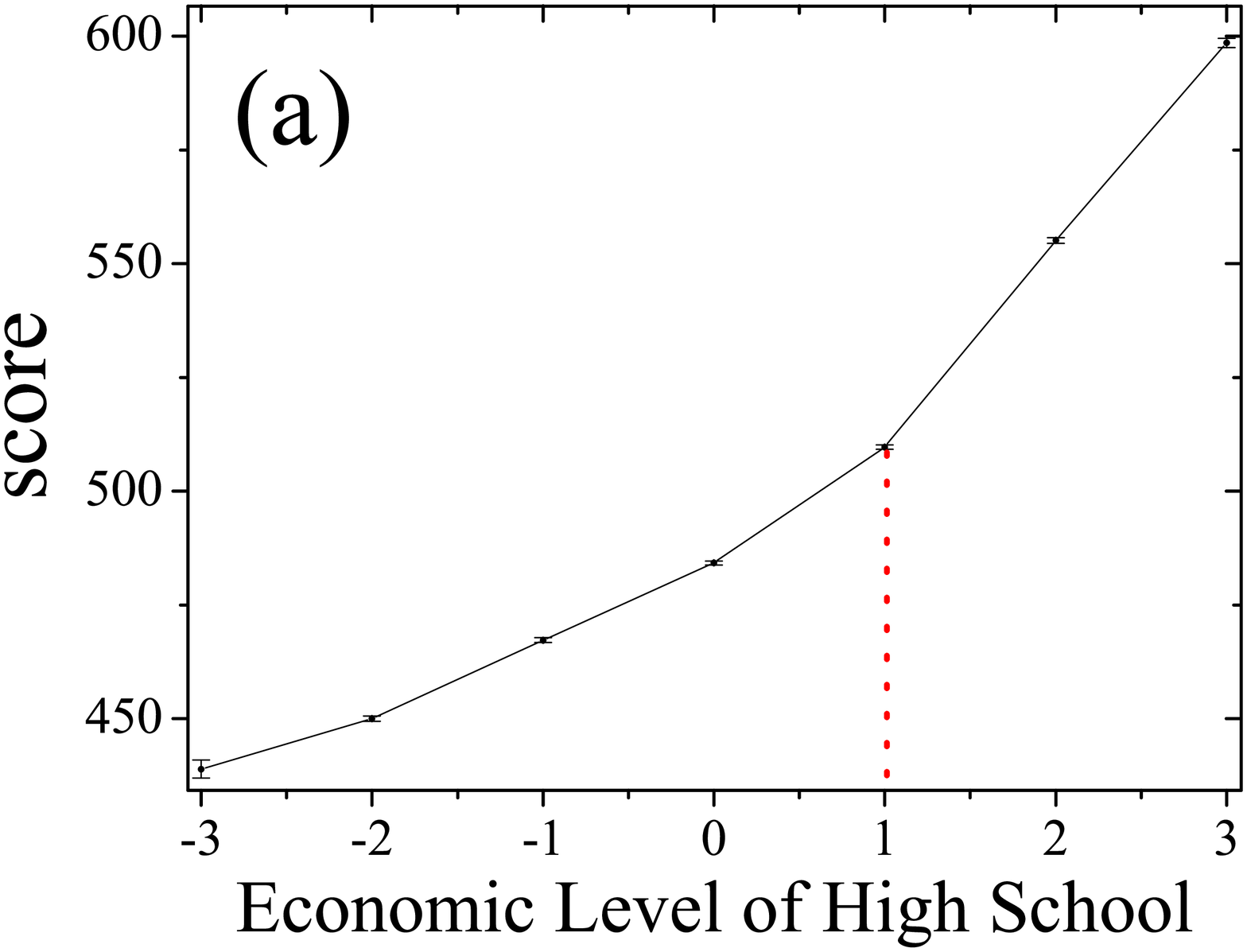} %
\includegraphics[width=0.75\columnwidth]{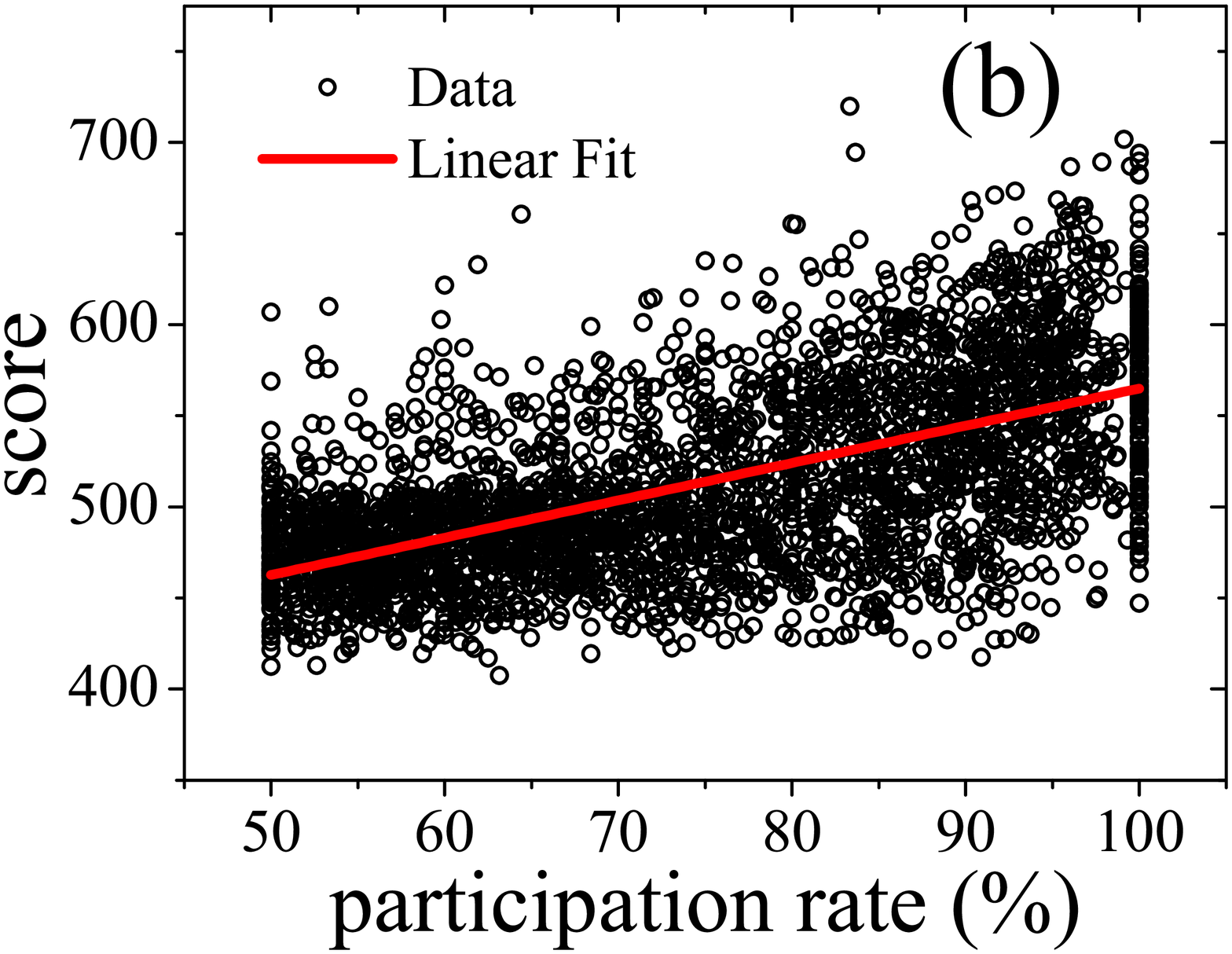}
\end{center}
\caption{(a) Averaged score as a function of the school social-economic
level from lower to upper level. (b) Score as function of students'
participation.}
\label{social_and_contingence}
\end{figure}

Figure~\ref{social_and_contingence} (a) shows the scores as a function of
the social-economic level of the high school. It is clear that the
social-economic level is quite relevant for the good performance of the
school. In particular, it is important to observe the large slope after the
``high average level''. The small error bars shows the reliable results. In
the Figure~\ref{social_and_contingence} (b) shows that the score increases
with the increase of the percent of the participation of the school, showing
a linear correlation namely $Score=360(2)+2.05(2)\cdot \rho $.

Another important test to check if the scores at different disciplines are
correlated is to compute the distribution of the scores. Here this
calculation is done in terms of the normalized value given by 
\begin{equation*}
z=\frac{\sqrt{n}(score-\left\langle score\right\rangle )}{\left\langle
\left( score-\left\langle score\right\rangle \right) ^{2}\right\rangle }%
\text{.}
\end{equation*}

\begin{figure}[th]
\begin{center}
\includegraphics[width=\columnwidth]{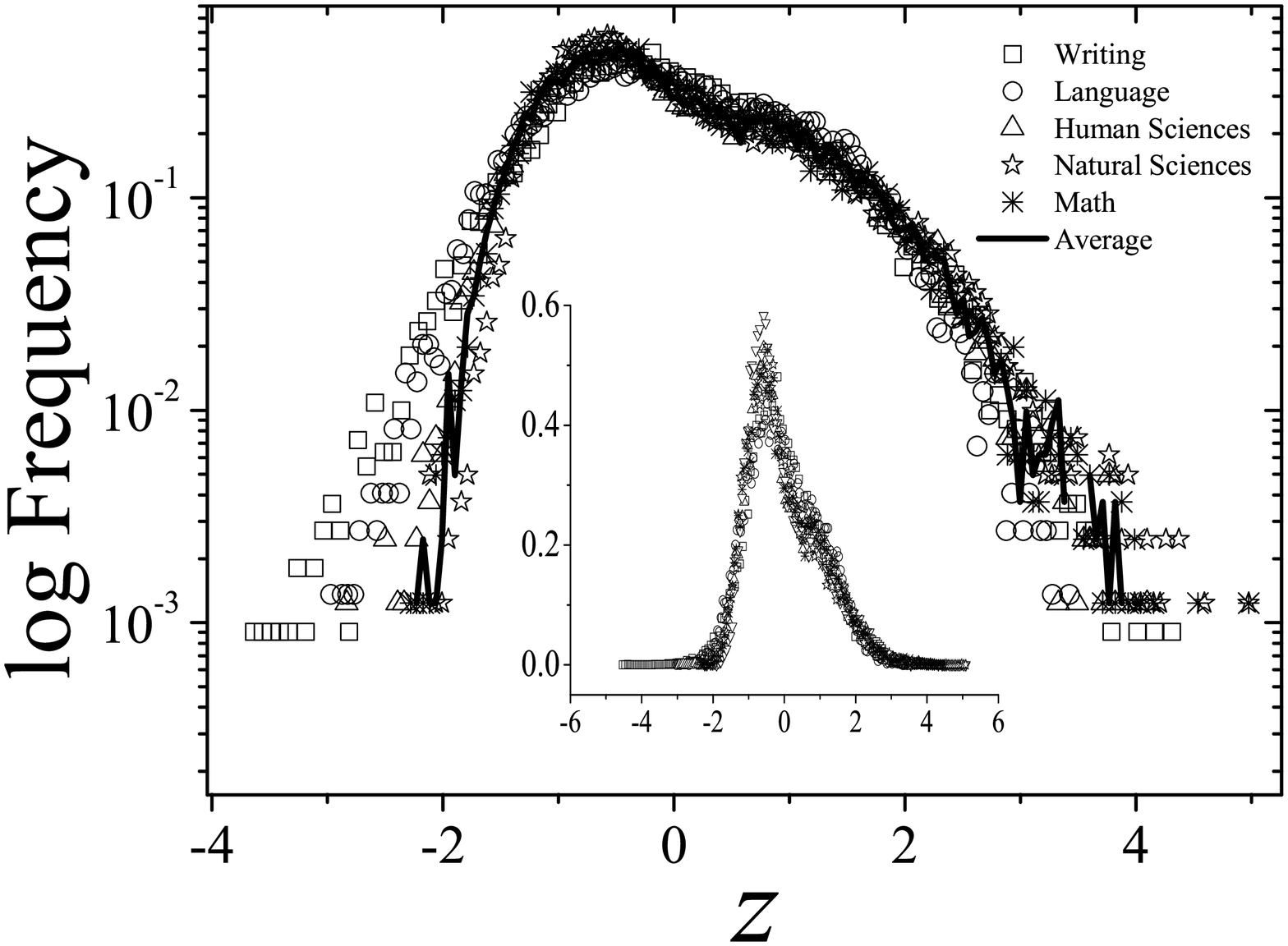}
\end{center}
\caption{Distributions of the scores for each subject in mono-log scale. All
the subjects follow in the same curve. The inset plot represents the same
data in a linear scale}
\label{universal}
\end{figure}

Figure~\ref{universal} shows the normalized scores distributions for the
different school subjects (points) in mono-log scale. The continuous curve
represents the average score distribution. The inset plot is depicted to
facilitate observation from the traditional linear scale point of view. It
is important to highlight that: all the different subjects obey the same
distribution of fluctuation of scores and this distribution is not Gaussian
since in mono-log we are not observing a second degree polynomial behavior.

What would be the distribution of the scores? In order to answer to this
questions, a few distributions shown in the Table~\ref{Table:distributions}
have been used to fit the scores of the schools. First, the standard
two-parametric statistical distributions (normal and log normal) were
checked. In this case $x_{c}$ and $\sigma $ were the free parameter for the
fit. Next, other more complicated asymmetrical distribution based on three
four parameters were also checked. 
\begin{figure}[th]
\begin{center}
\includegraphics[width=0.6\columnwidth]{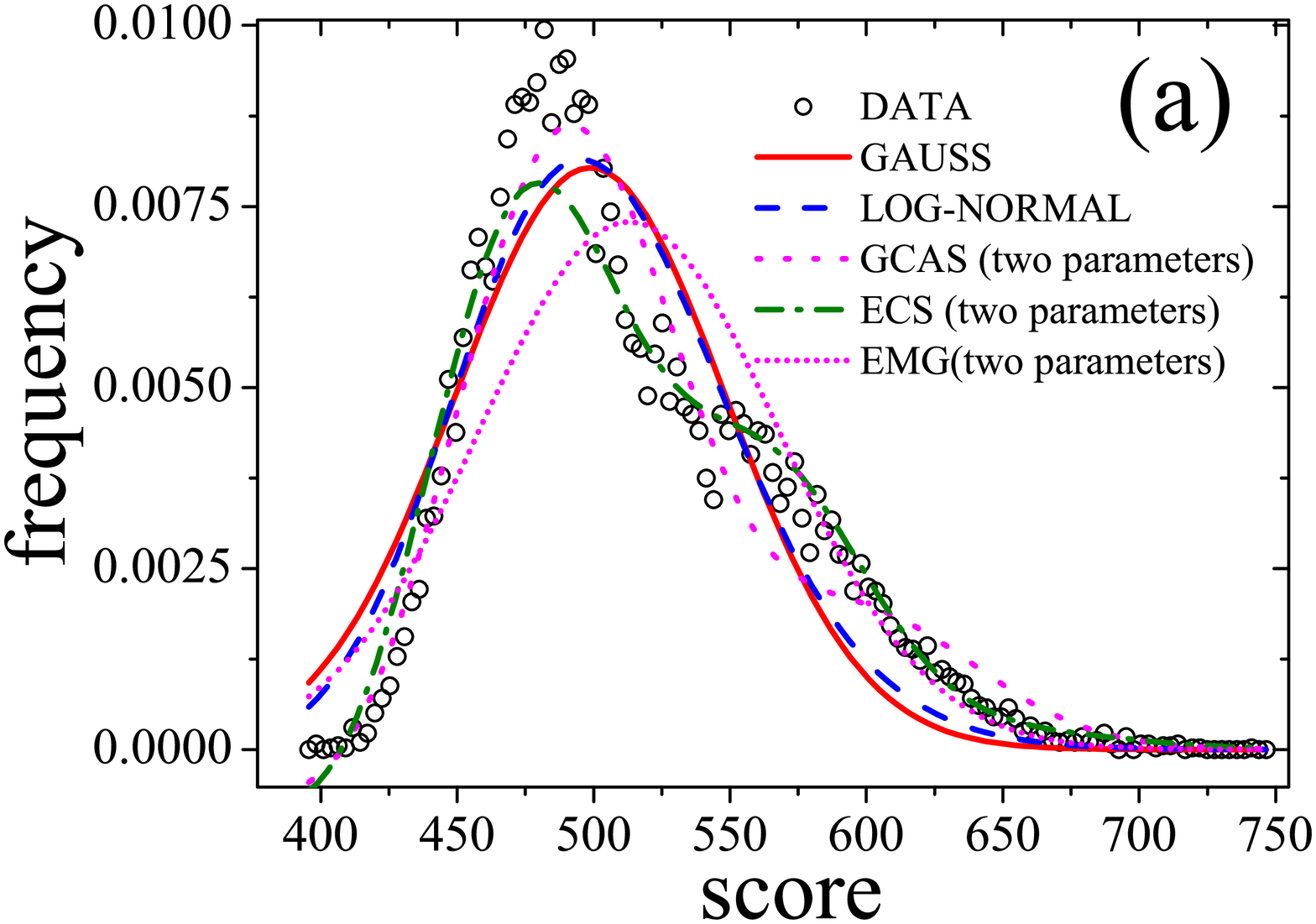}%
\includegraphics[width=0.6\columnwidth]{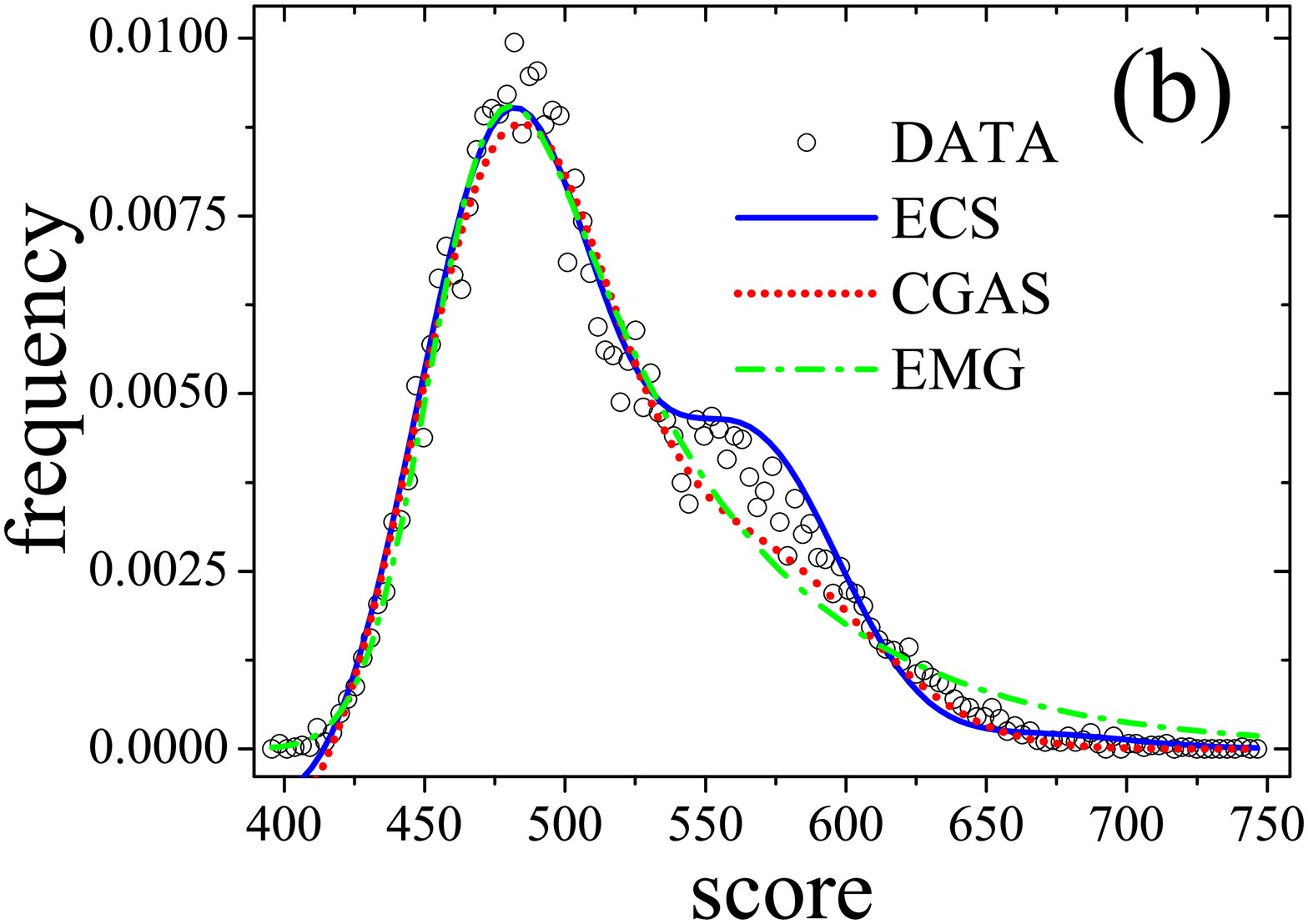}
\end{center}
\caption{Fits of the data using (a) two parameters distributions and
(b)three or four parameters distributions.}
\label{fits}
\end{figure}

The quality of the different fits performed here is checked by the following
procedure. Given the original data set $y_{1},...,y_{n}$ with $n$ values and
the fit of these values by the functions values $f_{1},..,f_{n}$ the
quantity of the fit is given by 
\begin{equation*}
R^{2}=1-\frac{SS_{res}}{SS_{tot}}=\frac{SS_{tot}-SS_{res}}{SS_{tot}}
\end{equation*}%
where $SS_{res}$, known as residual sum of squares is calculated by%
\begin{eqnarray}
SS_{res} &=&\sum_{i=1}^{n}(y_{i}-f_{i})^{2}  \notag \\
SS_{tot} &=&\sum_{i=1}^{n}(y_{i}-\overline{y})^{2}\;.  \label{Eq:SS}
\end{eqnarray}

In a general form, $R^{2}$ can be related to the unexplained variance, since
the second term compares the unexplained variance (variance of the model's
errors) with the total variance (of the data). Since $0\leq R^{2}\leq 1$,
with $R^2=1$ when original data and fit are identical, it gives a good
measurement of how far the fit is from the original data. It is also
important to mention that in the linear least squares regression, $R^{2}$ is
equal to the square of the Pearson correlation coefficient given by the Eq.~%
\ref{Eq.Pearson}.

Even thought the Gaussian distribution seems more promptly in nature, the
multiple parameters approach is seen in the movement of particles in random
media~\cite{Croma2012}, noise in semiconductor devices~\cite{noise2010},
stochastic aspects of soccer scores~\cite{soccer2013} which are described.
For all the tested distributions exemplified in the Table~\ref%
{Table:distributions} the parameter $R^{2}$ was computed. Here the tested
distributions are the normal or Gaussian (N), the log normal (LN), the
Exponentially Modified Gaussian (EMG),Gram-Charlier peak function (GC) and
Edgeworth-Cramer peak function (EC). Then, the fits using these
distributions were performed by the Levenberg-Marquardt method~\cite%
{recipes1992} for non-linear fits.

\begin{table}[tbp] \centering%
\begin{tabular}{lllll}
\hline\hline
\textbf{Dist.} & \textbf{Formula} & \textbf{parameters} & $R_{two}^{2}$ & $%
R_{all}^{2}$ \\ \hline\hline
N & 
\begin{tabular}{l}
$f(x)=\frac{e^{-(x-x_{c})^{2}/2\sigma ^{2}}}{\left( 2\pi \sigma ^{2}\right)
^{1/2}}$%
\end{tabular}
& 2: $x_{c}$ and\ $\sigma $ & $0.868$ & $0.868$ \\ 
LN & 
\begin{tabular}{l}
$f(x)=\frac{e^{-\frac{1}{2\sigma ^{2}}\ln ^{2}(x/x_{c})}}{\left( 2\pi \sigma
^{2}\right) ^{1/2}x}$%
\end{tabular}
& 2: $x_{c}$ and\ $\sigma $ & $0.899$ & $0.899$ \\ 
EMG & 
\begin{tabular}{l}
$f(x)=\frac{1}{t_{0}}e^{\frac{1}{2}(\frac{w}{t_{0}})^{2}-\frac{x-x_{c}}{t_{0}%
}}\int_{0}^{\frac{x-x_{c}}{w}-\frac{w}{t_{0}}}e^{-t^{2}}dt$%
\end{tabular}
& 3: $x_{c}$, $w$, and $t_{0}$ & $0.811$ & $0.971$ \\ 
GC & 
\begin{tabular}{l}
$f(x)=\frac{e^{-z^{2}/2}}{\left( 2\pi \sigma ^{2}\right) ^{1/2}}\left[ 1+%
\frac{a_{3}}{3!}(z^{3}-3z)+\frac{a_{4}}{4!}(z^{4}-6z^{3}+3)\right] $ \\ 
with $z=(x-x_{c})/\sigma $%
\end{tabular}
& 4: $x_{c}$, $\sigma $, $a_{3}$, and $a_{4}$ & $0.945$ & $0.970$ \\ 
EC & 
\begin{tabular}{l}
$f(x)=\frac{e^{-z^{2}/2}}{\left( 2\pi \sigma ^{2}\right) ^{1/2}}\left[ 1+%
\frac{b_{3}}{3!}(z^{3}-3z)+\frac{b_{4}}{4!}(z^{4}-6z^{3}+3)\right. $ \\ 
$\left. \frac{10b_{3}^{2}}{6!}(z^{6}-15z^{4}+45z-15)\right] $ with $%
z=(x-x_{c})/\sigma $%
\end{tabular}%
$\ $ & 5: $x_{c}$, $\sigma $, $b_{3}$, and $b_{4}$ & $0.958$ & $0.979$ \\ 
\hline\hline
\end{tabular}%
\caption{Functions used to fit the  distribution of scores, $x$, 
of  the schools. The last two columns show the determination 
coefficient $R^2$ by using, respectively, two and all parameters of the considered 
functions. For the computation of  $R$ for  EMG, GC, and EC 
with only two parameters, the parameters  $x_{c}$ and $\sigma$ were 
fixed by the average and standard deviation estimated from the original data}%
\label{Table:distributions}%
\end{table}%

In the case of the distributions EMG, GC, and EC, two approaches have been
employed. First, two parameters were estimated with a statistical measure
and the others fitted. For example in the case of the EMG, GC, and EC
distributions, $x_{c}$ was fixed with the average of the scores. This
procedure yields $x_c=\left\langle x \right\rangle =513.36$. Then, $\sigma $
is computed in the standard deviation of the original data. This gives $%
\sigma=\left\langle x^{2}\right\rangle -\left\langle x\right\rangle
^{2}=52.1 $. With $x_c$ and $\sigma$ fixed, the only free parameters for the
EMG, GC, and EC distributions become ($w$,$t_{0}$), ($a_{3}$,$a_{4}$), and ($%
b_{3},b_{4}$) respectively. In addition to the fit with two parameters, a
fit in which all the parameters was performed. The comparison between the
value of $R^2$ (see Eq.~\ref{Eq:R_square}) obtained using these two fitting
methods is illustrated in the Table~\ref{Table:distributions}.

The Figure~\ref{fits}(a) illustrates the comparison of the original data
with the N, LN, EMG,GC and EC, these three last employing a two parameters
fit. The visual inspection of the graphs support the results of the
determination coefficient $R^{2}$~\cite{Trivedi2002} shown in the Table~\ref%
{Table:distributions} that indicates that using two parameters the best
fitting distribution is the EC. In the Figure~\ref{fits} (b) the original
data is compared with the results for the distributions EMG, GC and EC but
using all the parameters for the fit. In this case the performance of the EC
is the more efficient and it is more efficient than when the adjustment is
done with only two parameters.

Even though the ENEM is constructed to give a standardized score of
individuals, this is not the case for the score of the schools. The
distribution shows a region with a peak at the score 500 and another peak at
the score 550 what presents two distinct score evolution. This observation
is supported by the Figure~\ref{social_and_contingence} which shows an
abrupt change in the slope of the averaged scores with the increase of
economic status of the school. It is important to point out that since the
schools' scores are not Gaussians, the schools' score evolution over time is
not a reliable measure since the score of one year can not be compared with
the score of the subsequent year, simply because it is not standardized.

\subsection{ENEM and UFRGS Students' Scores}

Next, the performance of the students is analyzed. In order to check how the
ENEM's selection differs from the traditional methods employed by the
Brazilian Universities in the past, we select to analyze the performance of
the students that have done both the ENEM and the entrance exam at the
UFRGS. It is important to emphasize that here we are comparing the
performance of the same people in both exams.

The table~\ref{Table:correlationsenem201020112012students} shows for the
years 2011, 2012 and 2013 the correlations, $r$, between the scores obtained
by the students in the different subjects at the ENEM tests.

\begin{table}[tbp] \centering%
\begin{tabular}{ll}
\textbf{2011} & $\left\{ 
\begin{tabular}{ccccc}
\hline\hline
Writing & Language & Human Sciences & Natural Sciences & Maths \\ 
\hline\hline
Writing & 0.349 & 0.343 & 0.313 & 0.232 \\ 
-- & Language & 0.710 & 0.668 & 0.599 \\ 
-- & -- & Human Sciences & \textbf{0.772} & \textbf{0.619} \\ 
-- & -- & -- & Natural Sciences & 0.723 \\ \hline\hline
\end{tabular}%
\right. $ \\ 
\textbf{2012} & $\left\{ 
\begin{tabular}{ccccc}
\hline\hline
Writing & Language & Human Sciences & Natural Sciences & Maths \\ 
\hline\hline
Writing & 0.362 & 0.360 & 0.345 & 0.261 \\ 
-- & Language & 0.744 & 0.673 & 0.575 \\ 
-- & -- & Human Sciences & \textbf{0.773} & \textbf{0.647} \\ 
-- & -- & -- & Natural Sciences & 0.725 \\ \hline\hline
\end{tabular}%
\right. $ \\ 
\textbf{2013} & $\left\{ 
\begin{tabular}{ccccc}
\hline\hline
Writing & Language & Human Sciences & Natural Sciences & Maths \\ 
\hline\hline
Writing & 0.463 & 0.477 & 0.445 & 0.378 \\ 
-- & Language & 0.769 & 0.675 & 0.597 \\ 
-- & -- & Human Sciences & \textbf{0.745} & \textbf{0.652} \\ 
-- & -- & -- & Natural Sciences & 0.766 \\ \hline\hline
\end{tabular}%
\right. $%
\end{tabular}%
\caption{Correlation coefficients, $r$,  between scores in
the different subjects at 
the  ENEM of the students that have also
done the entrance exam of UFRGS in the years
2011, 2012 and 2013.}\label{Table:correlationsenem201020112012students}%
\end{table}%

It is interesting to observe that the correlation between the students'
scores in all subjects is large with the exception of Writing. It is
particularly intriguing the high correlation between the scores on Human and
Natural Sciences and Math, usually topics at school in which the performance
of the students differs a lot. One possible explanation for this phenomena
is related to the fact that at the ENEM the questions are quite long with
the addition of a contextualization usually absent in the problem solve
texts in exact science. It is important to mention that the behavior is the
same for all the years we have analyzes in our work. The low correlation
between the Writing and the other topics can be understood because this is
the only part of the exam that is not manipulated by the standardized method.

In order to check if the high correlation between scores is a characteristic
of the standardized procedure employed at the ENEM or it is due to the
students' profile, the same analysis was performed for the score at the
entrance exam at UFRGS.

The Table~\ref{Table:correlationUFRGSvestibular} illustrates the correlation
between the students' scores at different subjects at the entrance at UFRGS
during the years of 2011, 2012 and 2013. The division in subject areas in
the UFRGS's exam is a little different from the ENEM's test. In the case of
UFRGS Natural Sciences is divide in Physics, Chemistry and Biology, while
Human Sciences is split in History and Geography. It is interesting to
notice that the correlation between Human Sciences and Natural Sciences is
much lower than the correlation observed in the ENEM and the clear high
correlation is present only between the Physics, Chemistry and Math as
traditionally is observed at the high schools. As in the case of ENEM,
writing haw a very small correlation with other topics. In the case of the
entrance at UFRGS, the Writing is not use for elimination but for
classification what means that his ability serves to discriminate between
people equality apt to enter the university which ones has the better skills
for communication.

Due to the difference between the correlations between topics in the two
exams becomes obvious the need to directly compare the scores in the same
topics. The Table~\ref{Table:correlationbetweenenemufrgs} illustrates this
comparison. The table shows that writing not only is not correlated with
other subjects within the same exam but also is not correlated with the
performance in other exams. In addition, the correlation between the scores
in other topics when the two exams are compared is not high with the
exception of Math.

\begin{table}[tbp] \centering%
\begin{tabular}{|l|l|}
\hline
2013 & $\left\{ 
\begin{tabular}{ccccccc}
\hline\hline
Writing & Math & Phys & Chem & Bio & Geo & Hist \\ \hline\hline
Writing & 0.381 & 0.327 & 0.366 & 0.369 & 0.319 & 0.372 \\ 
& Math & 0.744 & 0.731 & 0.652 & \textbf{0.576} & \textbf{0.583} \\ 
&  & Phys & 0.697 & 0.634 & \textbf{0.548} & \textbf{0.552} \\ 
&  &  & Chem & 0.671 & \textbf{0.559} & \textbf{0.587} \\ 
&  &  &  & Bio & \textbf{0.575} & \textbf{0.600} \\ 
&  &  &  &  & Geo & 0.587 \\ \hline\hline
\end{tabular}%
\right. $ \\ 
2012 & $\left\{ 
\begin{tabular}{ccccccc}
\hline\hline
Writing & Math & Phys & Chem & Bio & Geo & Hist \\ \hline\hline
Writing & 0.366 & 0.335 & 0.330 & 0.323 & 0.323 & 0.363 \\ 
& Math & 0.781 & 0.744 & 0.638 & \textbf{0.594} & \textbf{0.557} \\ 
&  & Phys & 0.750 & 0.663 & \textbf{0.585} & \textbf{0.570} \\ 
&  &  & Chem & 0.649 & \textbf{0.564} & \textbf{0.544} \\ 
&  &  &  & Bio & \textbf{0.534} & \textbf{0.543} \\ 
&  &  &  &  & Geo & 0.606 \\ \hline\hline
\end{tabular}%
\right. $ \\ 
2011 & $\left\{ 
\begin{tabular}{ccccccc}
\hline\hline
Writing & Math & Phys & Chem & Bio & Geo & Hist \\ \hline\hline
Writing & 0.307 & 0.308 & 0.319 & 0.322 & 0.314 & 0.319 \\ 
& Math & 0.736 & 0.732 & 0.611 & \textbf{0.587} & \textbf{0.524} \\ 
&  & Phys & 0.757 & 0.662 & \textbf{0.595} & \textbf{0.571} \\ 
&  &  & Chem & 0.687 & \textbf{0.608} & \textbf{0.556} \\ 
&  &  &  & Bio & \textbf{0.597} & \textbf{0.562} \\ 
&  &  &  &  & Geo & 0.602 \\ \hline\hline
\end{tabular}%
\right. $ \\ \hline
\end{tabular}%
\caption{Correlation, $r$,  between the scores of the
different subjects at  the UFRGS examinations in the
years 2011, 2012 and 2013}\label{Table:correlationUFRGSvestibular}%
\end{table}%
%
%
%
%

\begin{table}[tbp] \centering%
$%
\begin{tabular}{cccccccc}
\hline\hline
UFRGS-ENEM & Word & Math & Human/Geo & Human/Hist & Phys/Nat & Chem/Nat & 
Bio/Nat \\ \hline\hline
2011 & 0.313 & 0.700 & 0.627 & 0.628 & 0.643 & 0.668 & 0.653 \\ 
2012 & 0.313 & 0.728 & 0.654 & 0.687 & 0.684 & 0.676 & 0.641 \\ 
2013 & 0.384 & 0.759 & 0.613 & 0.673 & 0.679 & 0.692 & 0.681 \\ \hline\hline
\end{tabular}%
$%
\caption{Correlation between the scores 
at specific subjects, $r$, at the  ENEM and at the  UFRGS 
examinations in the years 2011, 2012 and 2013. }\label%
{Table:correlationbetweenenemufrgs}%
\end{table}%
%
%
%
%
%

Although the correlation is high, we would expect an ever higher correlation
between the two examinations if they intend to admit good candidates to the
university (UFRGS). Let us observe that the University has been able to
select good students and the institution has achieves high rankings in all
evaluations carried out by the Ministry of Education. UFRGS is consistently
ranked among the top 5 universities in Brazil for both research and
education. It is important to mention and note the small correlations
between the Writing test between two exams.

\begin{figure}[th]
\begin{center}
\includegraphics[width=0.5\columnwidth]{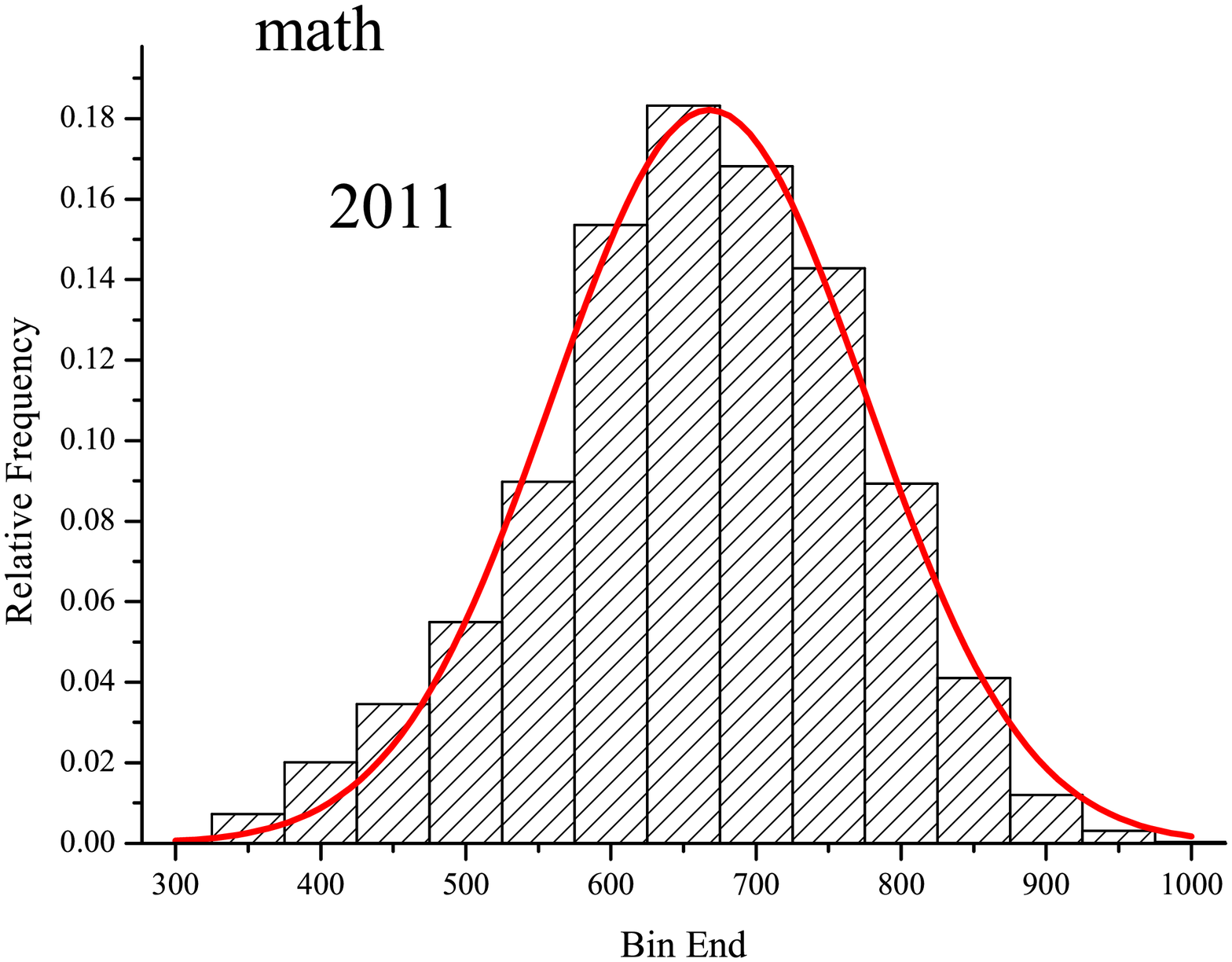}%
\includegraphics[width=0.5\columnwidth]{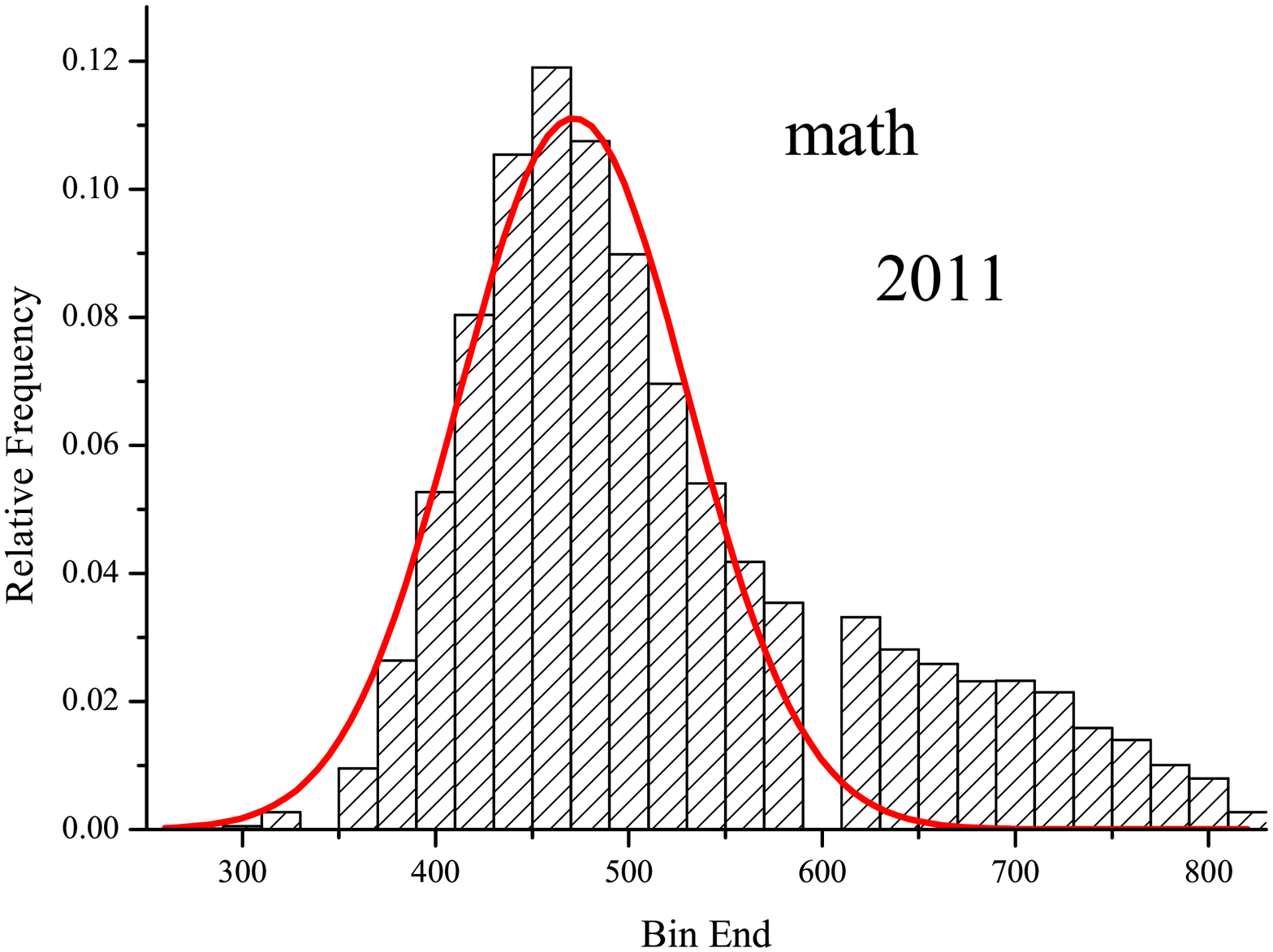} %
\includegraphics[width=0.5\columnwidth]{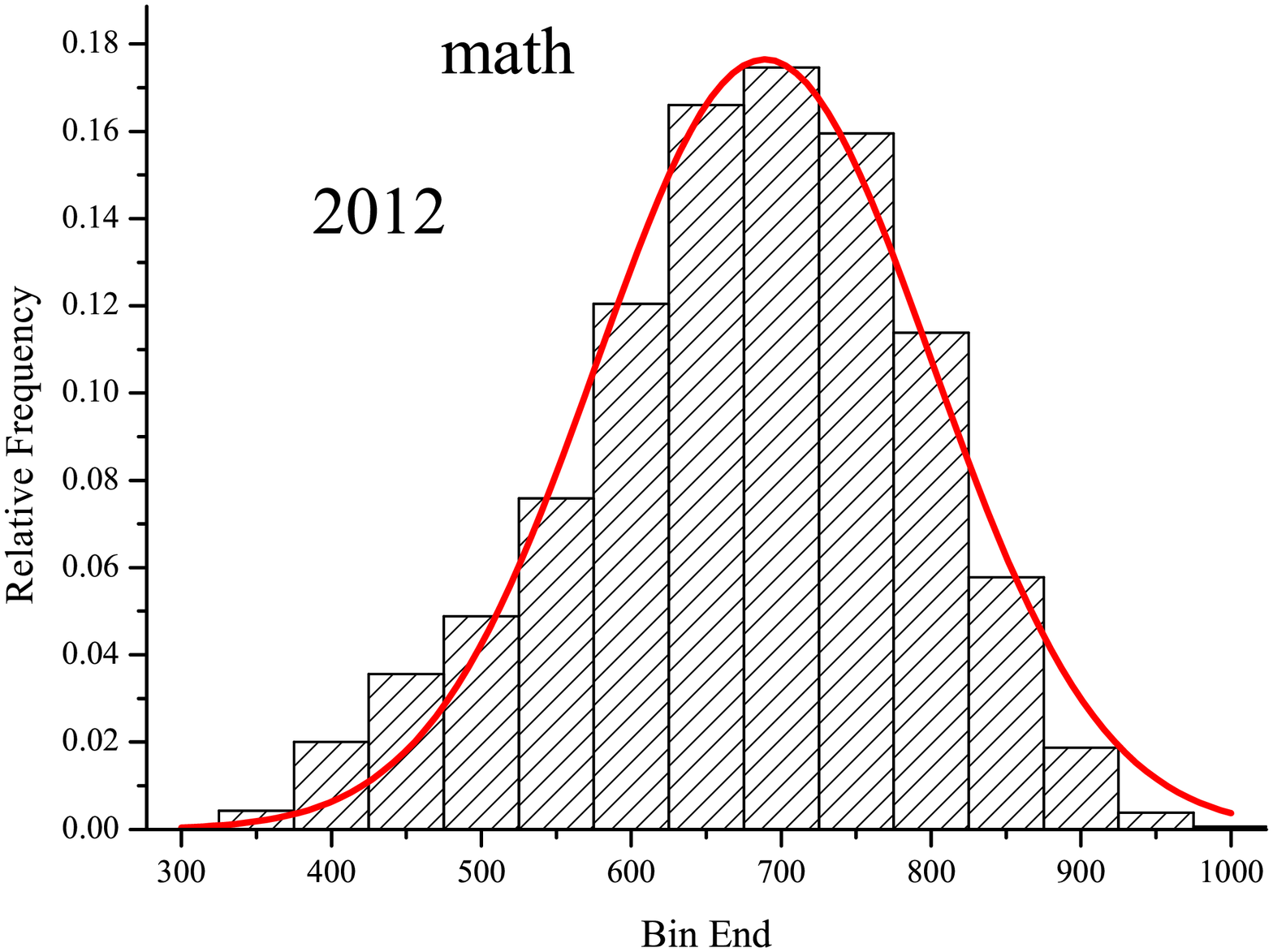}%
\includegraphics[width=0.5\columnwidth]{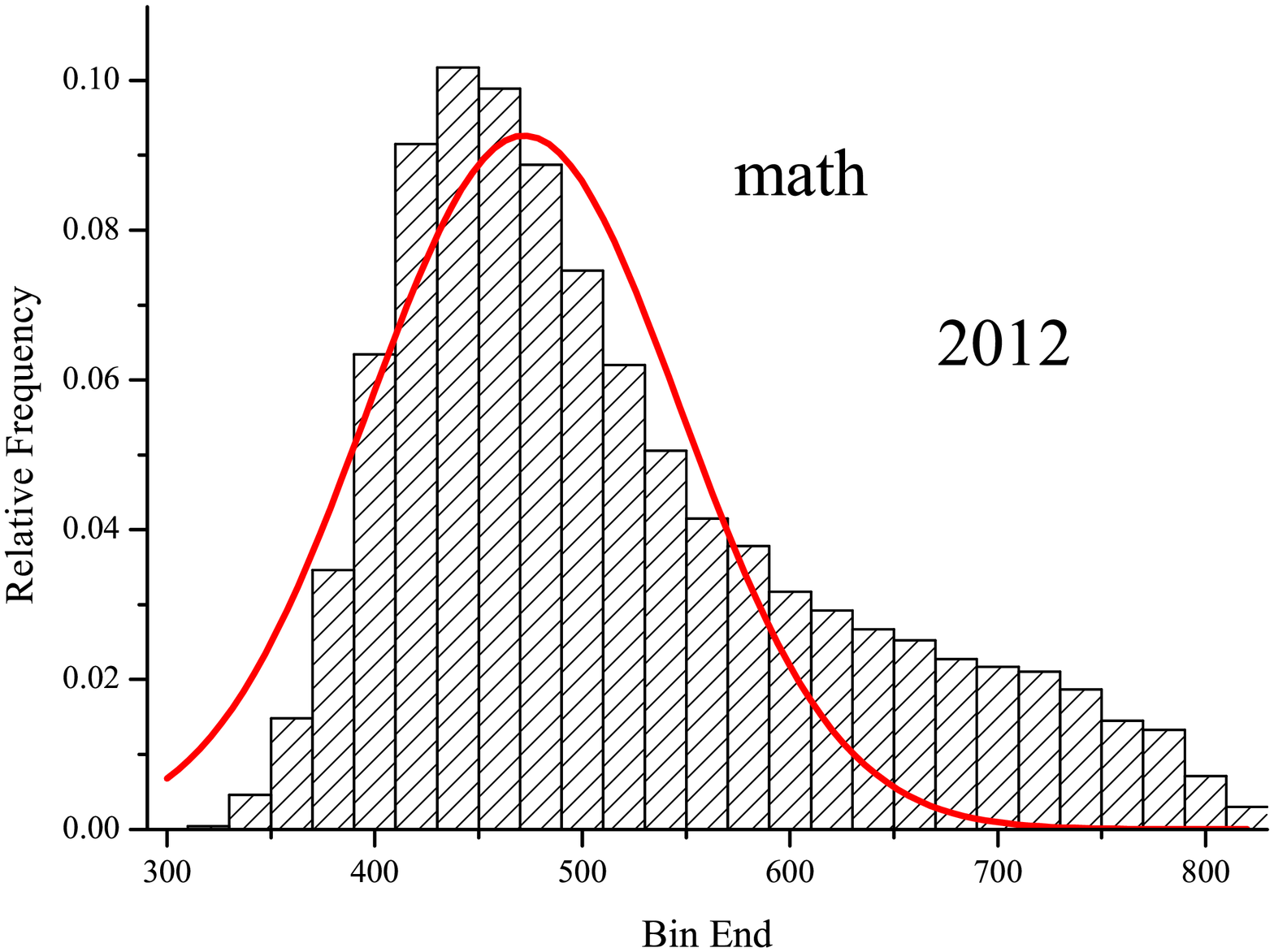} %
\includegraphics[width=0.5\columnwidth]{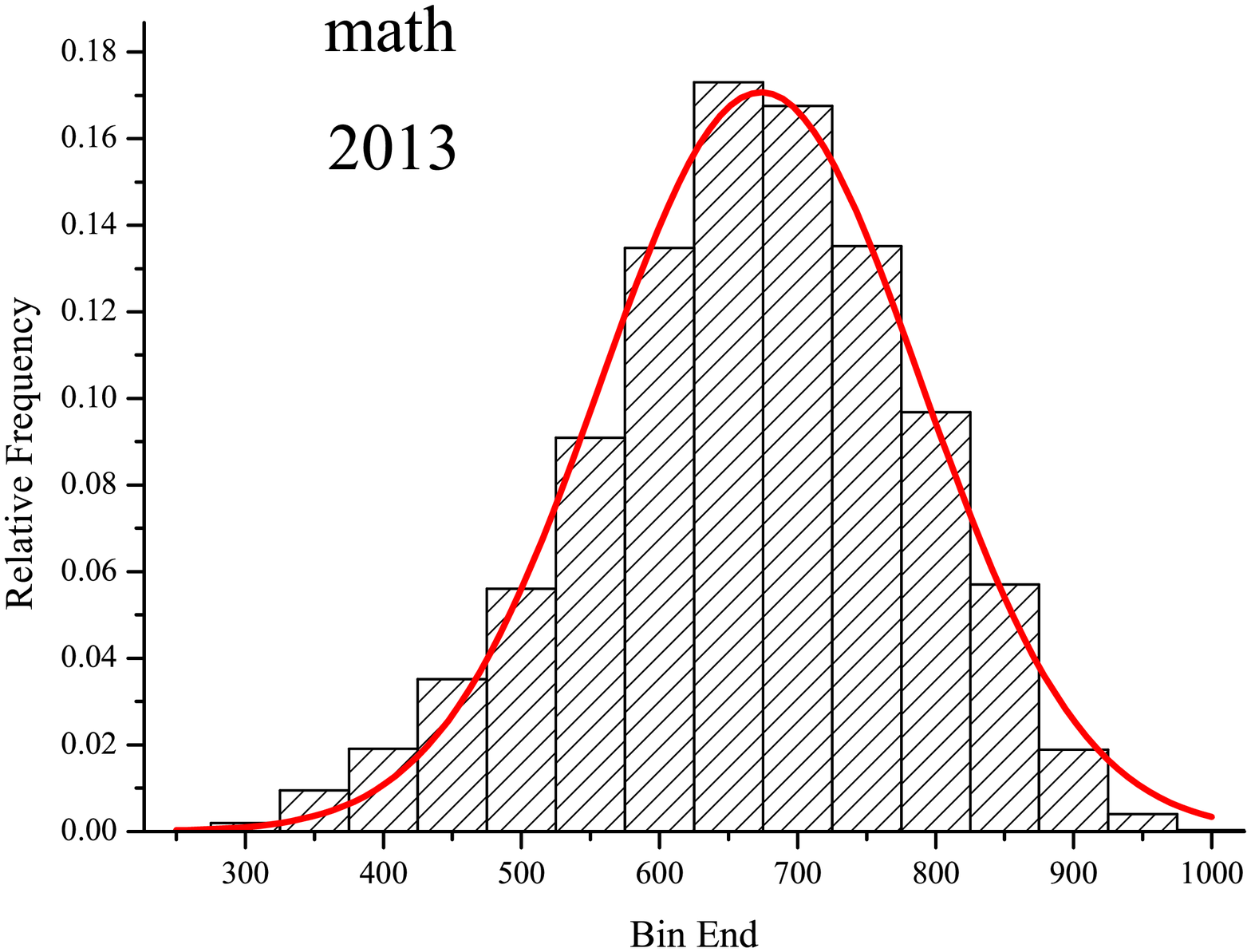}%
\includegraphics[width=0.5\columnwidth]{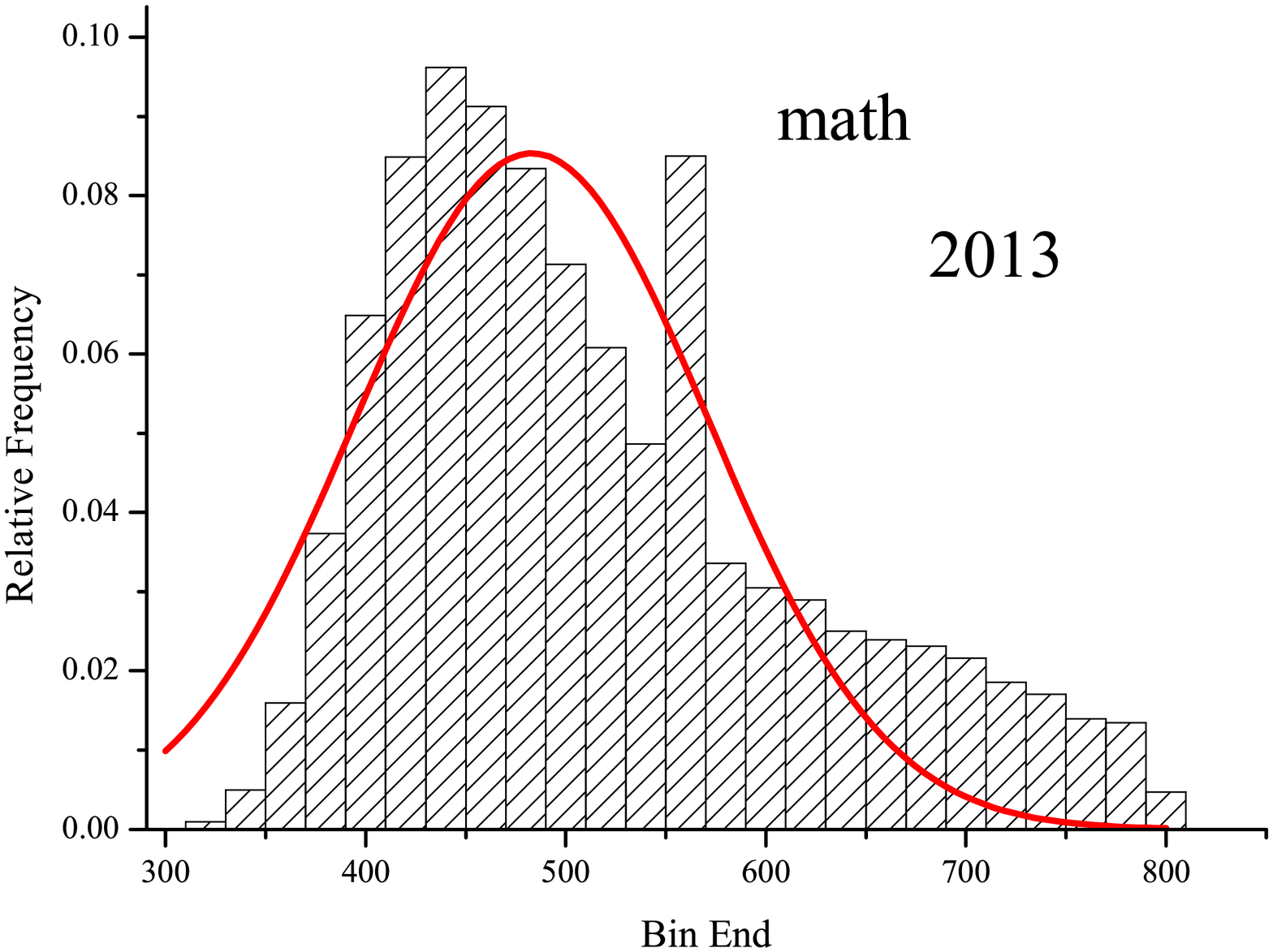}
\end{center}
\caption{Score distribution of the same candidates in UFRGS and ENEM for
Mathematics. The continuous curves correspond to Gaussian fits. We can
observe a deep difference in the right histograms (UFRGS) in comparison with
the left histograms (ENEM).}
\label{Fig:comparisomUFRGSENEMdistributions}
\end{figure}

The differences between the two exams was also checked by comparing the
distribution of the scores for Math. The Figure~\ref%
{Fig:comparisomUFRGSENEMdistributions} illustrates the ENEM's and the
UFRGS's distributions for Math for the years 2011, 2012 and 2013. The ENEM's
distributions are visually Gaussian forms while the UFRGS's exams show
distinctions when compared with the Gaussian. These similarities and
differences can be computed by two quantities: skewness and kurtosis.
Skewness is a measure of lack of symmetry. A distribution, or data set, is
symmetric if it looks the same to the left and right of the center point. A
symmetrical distribution has a skewness of zero, while an asymmetrical
distribution with right(left) tail has a positive(negative) skew. Kurtosis
is a measure of whether the data are peaked or flat relative to a normal
distribution. A Gaussian distribution has a kurtosis of 0, while a flatter
distribution has a negative kurtosis and a very peaked distribution has a
positive kurtosis. Table~\ref{Table:dist_enem} shows the kurtosis and the
skewness of the ENEM's score distributions in the years 2011, 2012 and 2013
while the Table~\ref{Table:dist_ufrgs} shows the kurtosis and the skewness
of the UFRGS's score distributions for the same period.

\begin{table}[tbp] \centering%
\begin{tabular}{llllllll}
\hline\hline
\textbf{Skewness} & \textbf{2011} & \textbf{2012} & \textbf{2013} & \textbf{%
Kurtosis} & \textbf{2011} & \textbf{2012} & \textbf{2013} \\ \hline\hline
Writing & $-0.056$ & $-8.7.10^{-4}$ & $+0.16$ & Writing & $-0.28$ & $-0.17$
& $-0.42$ \\ 
Language & $-0.54$ & $-0.60$ & $-0.35$ & Language & $+0.98$ & $+1.21$ & $%
+0.33$ \\ 
Humanities & $-0.46$ & $-0.24$ & $-0.31$ & Humanities & $+0.56$ & $+0.35$ & $%
+0.057$ \\ 
Natural Sciences & $-0.35$ & $-0.031$ & $-0.098$ & Natural Sciences & $+0.31$
& $+0.19$ & $-0.34$ \\ 
Math & $-0.26$ & $-0.36$ & $-0.29$ & Math & $-0.11$ & $-0.18$ & $-0.12$ \\ 
\hline\hline
\end{tabular}%
\caption{Skewness and kurtosis of the ENEM's score distributions in
the years 2011, 2012 and 2013.}\label{Table:dist_enem}%
\end{table}%
%

The tables show that there is a negative skewness for ENEM's Maths scores,
but positive in the UFRGS's scores in the analyzed years. The same occurs,
now shown here for simplicity, for Natural Sciences (ENEM) when compared
with Physics, Chemistry and Biology (UFRGS) and Humanities (ENEM) when
compared with History and Geography (UFRGS). For the kurtosis, for example,
we have opposite signals for the writing test in ENEM and UFRGS for Writing
and Humanities.

\begin{table}[tbp] \centering%
\begin{tabular}{llllllll}
\hline\hline
\textbf{Skewness} & \textbf{2011} & \textbf{2012} & \textbf{2013} & \textbf{%
Kurtosis} & \textbf{2011} & \textbf{2012} & \textbf{2013} \\ \hline\hline
Writing & $-0.19$ & $-0.33$ & $-0.17$ & Writing & $0.14$ & $0.64$ & $0.20$
\\ 
Geo & $0.34$ & $0.18$ & $0.15$ & Geo & $-0.28$ & $-0.44$ & $-0.24$ \\ 
Hist & $0.28$ & $0.25$ & $0.25$ & Hist & $-0.36$ & $-0.41$ & $-0.38$ \\ 
Math & $0.88$ & $0.82$ & $0.77$ & Math & $0.13$ & $-0.14$ & $-0.27$ \\ 
Phys & $0.85$ & $1.03$ & $0.90$ & Phys & $0.26$ & $0.58$ & $0.69$ \\ 
Chem & $0.89$ & $0.99$ & $0.81$ & Chem & $0.22$ & $0.59$ & $0.25$ \\ 
Bio & $0.64$ & $0.81$ & $0.63$ & Bio & $0.11$ & $0.62$ & $0.074$ \\ 
\hline\hline
\end{tabular}%
\caption{Skewness and kurtosis of the UFRGS's score distributions in
the years 2011, 2012 and 2013.}\label{Table:dist_ufrgs}%
\end{table}%
%
%
%
%
%
%
%
%
%
%
%
%
%
%

Such differences can be observed for a particular case, i.e. the Maths test.
We can see the deviation from normal of the UFRGS examination which is not
observed for the ENEM examinations. This corroborates the results found in
Tables \ref{Table:dist_enem} and \ref{Table:dist_ufrgs}.

This result suggests that the exams rank the students in a different order.
In order to check this hypothesis, the following strategy was employed. The
differences between the rankings of students according to their scores in
the two exams was obtained by denoting by $r_{ENEM}(i)$ the rank of the $i$%
-th student in the ENEM examinations and denoting by $r_{UFRGS}(i)$ the rank
corresponding to the same student at UFRGS. Then, the following quantity 
\begin{equation}
d_{i}=r_{UFRGS}(i)-r_{ENEM}(i)\text{,}  \label{Eq.d}
\end{equation}%
that measures the difference between the ranks established by the two exams
for a specific school subject was defined.

Then the average difference in the ranking $\alpha $ index becomes 
\begin{equation}
\alpha =\frac{\left\langle \left\vert d_{i}\right\vert \right\rangle }{%
N_{total}}=\frac{\sum_{i=1}^{N_{total}}\left\vert d_{i}\right\vert }{%
N_{total}^{2}}  \label{Eq.alfa}
\end{equation}%
where $N_{total}$ is the total number of analyzed students in which we
choose to represent it in percentages. It measures the average ranking
difference between the two exams. In the data, we excluded students with
score zero in one of analyzed exams for a fair comparison. In the Table \ref%
{Table:ranking_deviations} it is possible to observe the differences
determined by the two rankings considering two subjects, Maths and Writing.

\begin{table}[tbp] \centering%
\begin{tabular}{c}
\hline\hline
\textbf{Math} \\ \hline\hline
\multicolumn{1}{l}{%
\begin{tabular}{|l|c|c|c|c|c|c|c|}
\hline\hline
\textbf{year} & $\left\langle \left\vert d_{i}\right\vert \right\rangle $ & $%
\left\langle d_{i}^{2}\right\rangle -\left\langle d_{i}\right\rangle ^{2}$ & 
$\left\langle d_{i}^{2}\right\rangle -\left\langle \left\vert
d_{i}\right\vert \right\rangle ^{2}$ & $\max_{i}\left\vert d_{i}\right\vert $
& $N_{total}$ & $\alpha $ & $\beta $ \\ \hline\hline
2011 & $3641$ & $4758$ & $3063$ & $19549$ & $21510$ & $16.9\%$ & $%
\allowbreak 17\%$ \\ 
2012 & $3501$ & $4589$ & $2964$ & $20267$ & $22651$ & $15.4\%$ & $15\%$ \\ 
2013 & $3628$ & $4792$ & $3131$ & $20532$ & $25023$ & $14.5\%$ & $14\%\,$ \\ 
\hline\hline
\end{tabular}%
} \\ \hline\hline
\textbf{Writing} \\ \hline\hline
\multicolumn{1}{l}{%
\begin{tabular}{|l|c|c|c|c|c|c|c|}
\hline\hline
\textbf{year} & $\left\langle \left\vert d_{i}\right\vert \right\rangle $ & $%
\left\langle d_{i}^{2}\right\rangle -\left\langle d_{i}\right\rangle ^{2}$ & 
$\left\langle d_{i}^{2}\right\rangle -\left\langle \left\vert
d_{i}\right\vert \right\rangle ^{2}$ & $\max_{i}\left\vert d_{i}\right\vert $
& $N_{total}$ & $\alpha $ & $\beta $ \\ \hline\hline
2011 & $2834$ & $3581$ & $2188$ & $10315$ & $10559$ & $26.8\%$ & $33\%$ \\ 
2012 & $2922$ & $3685$ & $2245$ & $10761$ & $10857$ & $26.9\%$ & $37\%$ \\ 
2013 & $3156$ & $4010$ & $2472$ & $11868$ & $12423$ & $25.4\%$ & $34\%$ \\ 
\hline\hline
\end{tabular}%
}%
\end{tabular}%
\caption{Statistics about ranking deviation  between ENEM and UFRGS in Maths
and Writing}\label{Table:ranking_deviations}%
\end{table}%

\begin{figure}[th]
\begin{center}
\includegraphics[width=\columnwidth]{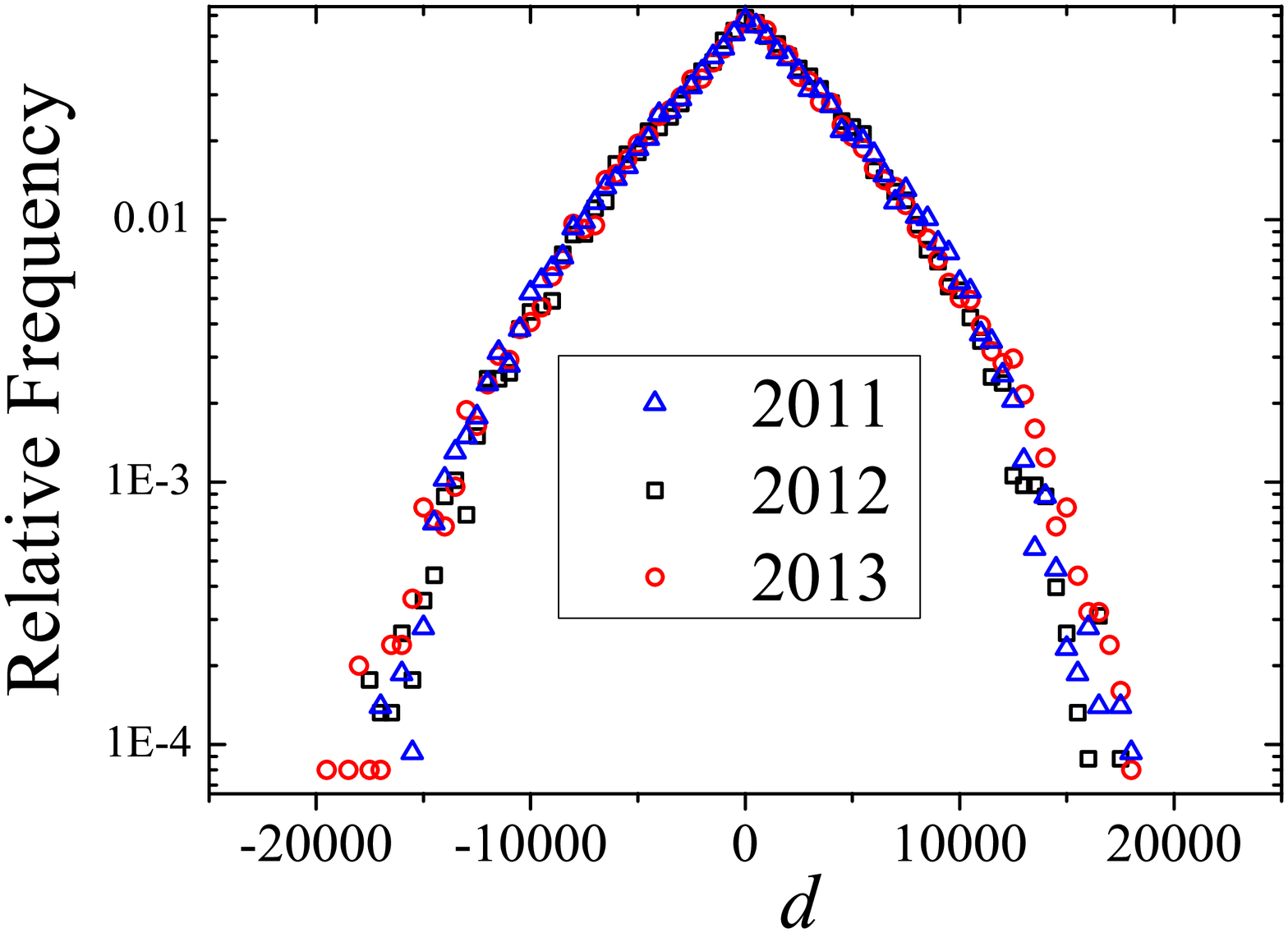}
\end{center}
\caption{Histogram of rank differences between UFRGS and ENEM for Maths in
mono log scale. A universality is observed under different years analyzed. }
\label{Fig:rank_difference}
\end{figure}

It is important to observe that the ranking of ENEM does not match that of
UFRGS. We can observe that a maximum difference $\max_{i}\left\vert
d_{i}\right\vert $ (in the Table \ref{Table:ranking_deviations}) is near to
the maximum possible difference ($N_{total}$). The plot of the histogram of
the rank differences, i.e., $d_{i}$, $i=1,...,N_{total}$, can be observed in
the Figure~\ref{Fig:rank_difference}.

Although the differences $d_{i}$ are distributed around zero, we can observe
that the standard deviation of $\left\vert d_{i}\right\vert $ is very large
according to \ref{Table:ranking_deviations}. The average difference in
Maths, considering the three years for example is around $3,550$ positions
which is a very large difference when one considers that ENEM will be used
as a national exam. In order to understand the coefficient $\alpha $ we
performed a simple numerical simulations. Basically we consider $N_{total}$
numbers in ascending order. We build from this ordered list a partially
randomized list by performing $\left\lceil \beta N_{total}\right\rceil $
swaps between pairs of numbers randomly chosen and independently on their
positions. Now with this new list in hands we calculate $\left\langle
\left\vert d_{i}\right\vert \right\rangle _{\text{rand}}$. The optimization
method concerts to find the best $\beta $ such that $\left\langle \left\vert
d_{i}\right\vert \right\rangle _{\text{rand}}$ is nearest the $\left\langle
\left\vert d_{i}\right\vert \right\rangle _{\text{real}}$ corresponding to
the ranking obtained by the experimental data between two exams (second
columns in Table \ref{Table:ranking_deviations}). A pseudo-code of algorithm
used to find the optimal $\beta $, which we so called Optimal\_Beta, can be
checked in Table \ref{Table:procedure_calcula_beta}. In this algorithm rand$%
(idum)$ is a pseudo-random number and $idum$ is the respective seed used to
generate the sequence of these numbers. The symbol {*}/ denotes the comments
of pseudo-code.

\begin{table}[tbp] \centering%
\begin{tabular}{ll}
\hline\hline
& \textbf{Procedure Optimal\_Beta} ($\beta _{\text{min}},\beta _{\text{max}%
},N_{total},$ $\left\langle \left\vert d_{i}\right\vert \right\rangle _{%
\text{real}},\Delta \beta $) \\ \hline\hline
& \textbf{input}: $\beta _{\text{min}},\beta _{\text{max}},N_{total},$ $%
\left\langle \left\vert d_{i}\right\vert \right\rangle _{\text{real}}$ \\ 
& \textbf{output}: $\beta _{opt}$ \\ 
& \textbf{Vector}: $v[i=1,...,N_{total}]$ \\ 
{*}/ & \textbf{Initializations}: \\ 
& $\Delta =N_{total}^{2}$ (or other big number of your choice) \\ 
& For $i=1,...,N_{total}$ \\ 
& \ \ $v[i]=i$ \\ 
& Endfor \\ 
{*}/ & Span $\beta $ from $\beta _{\text{min}}$ up to $\beta _{\text{max}}$
with precision $\Delta \beta $: \\ 
& For $\beta =\beta _{\text{min}},\beta _{\text{max}};\Delta \beta $ \\ 
& For $icount=1,\left\lceil \beta N_{total}\right\rceil $ \\ 
& \ \ \ \ \ $i:=\ $rand$(idum)\ast N_{total}+1$ \\ 
& $\ \ \ \ \ j:=\ $rand$(idum)\ast N_{total}+1$ \\ 
{*}/ & Perform the swap! \\ 
& $\ \ \ \ \ aux:=v(i)$ \\ 
& $\ \ \ \ \ v(i):=v(j)$ \\ 
& $\ \ \ \ \ v(j)=aux$ \\ 
& EndFor \\ 
{*}/ & Compute $\left\langle \left\vert d_{i}\right\vert \right\rangle _{%
\text{rand}}$, i.e, the average distance between the \\ 
{*}/ & randomized list and ordered $N_{total}$ numbers; \\ 
& \ For $i=1,N_{total}$ \\ 
& $\ \ \ \left\langle \left\vert d_{i}\right\vert \right\rangle _{\text{rand}%
}=\left\langle \left\vert d_{i}\right\vert \right\rangle _{\text{rand}%
}+\left\vert i-v(i)\right\vert $ \\ 
& \ Endfor \\ 
& $\ \ \ \ \left\langle \left\vert d_{i}\right\vert \right\rangle _{\text{%
rand}}=\left\langle \left\vert d_{i}\right\vert \right\rangle _{\text{rand}%
}/N_{total}$ \\ 
& $\ \ \ \ temp:=\left\vert \left\langle \left\vert d_{i}\right\vert
\right\rangle _{\text{rand}}-\left\langle \left\vert d_{i}\right\vert
\right\rangle _{\text{real}}\right\vert $ \\ 
& \ \ \ \ \ \ \ \ \ If ($temp<\Delta $) then \\ 
& $\ \ \ \ \ \ \ \ \ \ \ \ \ \beta _{opt}:=\beta ;$ \\ 
& $\ \ \ \ \ \ \ \ \ \ \ \ \Delta :=temp;$ \\ 
& \ \ \ \ \ \ \ \ \ Endif \\ 
& Endfor \\ 
& Return $\beta _{opt}$ \\ 
& Stop \\ 
& End \\ \hline\hline
\end{tabular}%
\caption{Procedure for computing the $\beta$ index}\label%
{Table:procedure_calcula_beta}%
\end{table}%

The $\beta -$values are shown in last columns in table \ref%
{Table:ranking_deviations}. There are a notorious correspondence between $%
\alpha $ and $\beta $ which corroborates the definition used to measure the
difference between two rankings.

We are convinced that all factors previously raised with respect to ENEM,
such as the size of question statements, the duration of the exam provides
conditions to less prepared students to obtain similar scores of well
prepared students that have more comprehensive knowledge. This is observed
by the statistics related to score distribution: such statistics show an
apparent homogenization of the evaluation system process when actually it
should separate the well-prepared and the other candidates.

\section{Conclusions}

\label{Section:Conclusions}

Standardized university entrance exams have been employed in many countries.
They share the characteristics of formatting the distribution of scores to
be fitted by a Gaussian. In this paper we study one particular standardized
test, the Brazilian's ENEM examination.

We found that unlike the students' scores distribution, the schools' scores
do not follow the Gaussian, but forms a two peaked distribution best fitted
by an EC distribution. This reflects the fact that the average schools'
score increases linearly with the economic level of the school in two
distinct regions with different slopes. This indicates that the exam is
designed to identify skills that are more commonly found in the economic
elite of the country. One possible explanation is the that since the exam is
very long, it requires that the students would be trained to spend hours
focusing on one specific topic, which is a kind of training that the more
expensive schools are able to provide.

Since the schools' scores distribution is not a Gaussian, it cannot be used
to compare the schools' performance over time since it is not a standardized
measure.

Next, the score of the students in the ENEM and in the UFRGS exams were
compared. The correlation between different subject in both cases can be
observed. Surprisingly, the correlation between Human Sciences and Natural
Sciences and Math is quite high in the case of the ENEM, which suggests that
the exam is not measuring the specific abilities in the different themes.

Since the ENEM's scores and the UFRGS's scores follow very different
distributions the change from one standardized test to a more itemize exam
implies selecting a different type of student. In summary, we employed
statistical methods to understand the characteristics of the selection in
two exams: one standardized test and a non standardized exam. Our results
indicate that there are differences in the selection of students is obtained
when each one of these exams is performed. It would be interesting in the
future to compared through the efficiency of higher education, ENADE, the
performance as professionals of the students selected by each one of these
methods~\cite{Zoghbi2012,Gupta2002}.

\section*{Acknowledgment}

We thank the Brazilian agencies CNPq, INCT-FCx, and Capes for the financial
support. We also thank the fruitful discussions with prof. Fernando Lang da
Silveira.


\end{document}